\documentclass[preprint,prd,aps,showpacs,showkeys,nofootinbib]{revtex4}
\usepackage{graphicx}
\usepackage{dcolumn}
\usepackage{bm}
\begin{document}

\preprint{hep-ph/0412147}

\title{Two-loop gluino contributions to neutron electric
dipole moment in CP violating MSSM}

\author{Tai-Fu Feng$^a$, Xue-Qian Li$^{b,c}$, Jukka Maalampi$^a$,
Xinmin Zhang$^c$}
\affiliation{$^a$ Department of Physics, 40014 University of Jyv\"askyl\"a,Finland}

\affiliation{$^b$ Department of Physics, Nankai University, Tianjing 300071, P. R. China}

\affiliation{$^c$ Institute of High Energy Physics, Academy of Science of China, 
P.O. Box 918, Beijing 100039, P. R. China}


\date{\today}

\begin{abstract}
We analyze two-loop gluino corrections to the neutron electric
dipole moment (EDM) in the minimal supersymmetry extension of the
standard model (MSSM). The dependence of two-loop corrections on
the relevant CP violating phases differs from that of the
one-loop contributions, and there is a region in the parameter space where the
two-loop contributions are comparable with the one-loop
contributions. Our numerical results show that the two-loop
corrections can be as large as 30\% of the one-loop results.
\end{abstract}



\maketitle

\section{Introduction}
\indent\indent The fermion electric dipole moments (EDMs) offer a powerful probe
for new physics beyond the Standard Model (SM). In the SM, the EDM
of the neutron is fully induced by the CP phase of the
Cabibbo-Kobayashi-Maskawa (CKM) matrix elements and it is predicted to be much
smaller \cite{smt} than the present experimental upper limit of
$1.1 \times10^{-25}\;e\cdot cm$ \cite{exp} and beyond the reach of
experiment in the near future. As for the minimal
supersymmetric extension of the SM (MSSM), there are many new
sources of the CP violation that can result in larger
contributions to the EDM of the neutron \cite{cp1,cp2}. Taking the
CP phases with a natural size of ${\cal O}(1)$, and the
supersymmetry mass spectra at the TeV range, the theoretical
prediction on the neutron EDM at one-loop level already exceeds
the present experimental upper bound. In order to make the
theoretical prediction consistent with the experimental
data, three approaches are adopted in the literature. One possibility is
to make the CP phases sufficiently small, i.e. $\le 10^{-2}$
\cite{cp1}. One can also assume a mass
suppression by making the supersymmetry spectra heavy, i.e. in the
several TeV range \cite{cp2}, or invoke a cancellation
among the different contributions to the fermion EDMs \cite{cp3}.

The prediction for the fermion
EDMs at one-loop level in a supersymmetric theory has been extensively discussed 
in the literature. On other hand,
the analysis on the neutron EDM at two-loop level is less
advanced, even though some pioneer work has been carried
out, for example the two-loop
Barr-Zee type diagrams involving the Higgs bosons \cite{cp4} and
the purely gluonic dimension-six Weinberg operator induced by
two-loop gluino-squark diagrams \cite{cp5} have been analyzed. 

Analyzing the EDMs at two-loop order can give us a better
understanding of where the new physics scale may emerge and shed some light to the
spectra of new physical particles  around this scale.
Moreover, the two-loop analysis involves some new parameters
in addition to those appearing in the one-loop calculations, hence resulting in 
more rigorous constraints on the supersymmetry parameter
space.

In this work we shall analyse the two-loop gluino
corrections to the neutron EDM in the MSSM. We work in the
framework of the simplest model, where we will neglect all other possible sources 
of
flavor violation except those related to the CKM matrix, and try to avoid
ambiguities of the unification conditions of the soft-breaking
parameters at the grand unification scale present in the 
mSUGRA schemes. In the next Section, we will demonstrate how to obtain the 
two-loop
gluino corrections to the neutron EDM. In the Section
\ref{sec3} we will present the results of our
numerical computations and we study the dependence of the
neutron EDM on the supersymmetry parameters. Conclusions are presented in  Section 
\ref{sec4}.

\section{The two-loop gluino corrections to the neutron EDM \label{sec2}}
\indent\indent
In the effective Lagrangian, the fermion EDM $d_{_f}$ is defined 
through the dimension five operator
\begin{eqnarray}
&&{\cal L}_{_{EDM}}=-{i\over2}d_{_f}\overline{f}\sigma^{\mu\nu}\gamma_5
fF_{_{\mu\nu}}
\label{eq1}
\end{eqnarray}
where$f$ is a fermion field,  $F_{_{\mu\nu}}$ is the 
electromagnetic field strength. This
coupling obviously signifies a CP violation. It is not present
among the fundamental interactions at tree level, but it is generated in loop 
level in an electroweak theory with CP violation.
Moreover, because quarks also take part in strong interactions, the
chromoelectric dipole moment (CEDM)
$\overline{f}T^a\sigma^{\mu\nu}\gamma_5 fG^a_{_{\mu\nu}}$ of
quarks, where $T^a\;(a=1,\;\cdots,\;8)$ denote the generators of
the strong $SU(3)$ gauge group, $G^a_{_{\mu\nu}}$ is the gluon field strength, and 
the pure gluon  Weinberg
operator of dimension-six,
$f_{_{abc}}G_{_{\mu\rho}}^aG^{b\rho}_{_\nu}
G_{_{\lambda\sigma}}^c\epsilon^{\mu\nu\lambda\sigma}$, contributes to the quark 
EDMs as well.  

A convenient way to describe 
loop-induced contributions is the effective theory approach, where
the heavy particles are integrated out at the matching scale. The resulting 
effective Lagrangian  includes a full set of CP violation
operators. In this work, we restrict ourselves to the following
operators that are relevant to the neutron EDM:
\begin{eqnarray}
&&{\cal L}_{_{eff}}=\sum\limits_{i}^5C_{_i}(\Lambda){\cal Q}_{_i}(\Lambda)\;,
\label{eq2}
\end{eqnarray}
where $C_{_i}(\Lambda)$ are the Wilson coefficients evaluated at the
scale $\Lambda$, and the five operators of interests are
\begin{eqnarray}
&&{\cal Q}_{_1}=\overline{q}\sigma^{\mu\nu}\omega_-qF_{_{\mu\nu}}
\;,\nonumber\\
&&{\cal Q}_{_2}=\overline{q}\sigma^{\mu\nu}\omega_+qF_{_{\mu\nu}}
\;,\nonumber\\
&&{\cal Q}_{_3}=\overline{q}T^a\sigma^{\mu\nu}\omega_-qG^a_{_{\mu\nu}}
\;,\nonumber\\
&&{\cal Q}_{_4}=\overline{q}T^a\sigma^{\mu\nu}\omega_+qG^a_{_{\mu\nu}}
\;,\nonumber\\
&&{\cal Q}_{_5}=-{1\over6}f_{_{abc}}G_{_{\mu\rho}}^aG^{b\rho}_{_\nu}
G_{_{\lambda\sigma}}^c\epsilon^{\mu\nu\lambda\sigma}\;.
\label{eq3}
\end{eqnarray}

\begin{figure}[t]
\setlength{\unitlength}{1mm}
\begin{center}
\begin{picture}(0,20)(0,0)
\put(-62,-100){\includegraphics{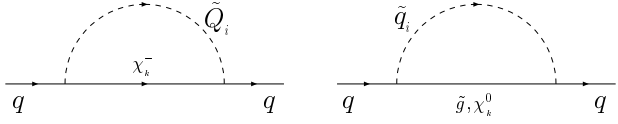}}
\end{picture}
\caption[]{The one-loop self energy diagrams which lead to the
quark EDMs and CEDMs in  MSSM, the corresponding triangle diagrams
are obtained by attaching a photon or a gluon in all possible ways
to the SUSY particles.} \label{fig1}
\end{center}
\end{figure}

The effective triangle diagrams 
responsible for the quark EDMs and CEDMs, are obtained by
attaching a photon or gluon external line,for quark EDM or CEDM,
respectively, in all possible ways to the quark self-energy
diagrams.

The one-loop supersymmetric corrections to the Wilson coefficients $C_i(\Lambda)$
in Eq. (\ref{eq2}) originate from three types of the graphs: the
quark self-energy diagrams, where one has a gluino and a squark,
a chargino and a squark or a neutralino and a squark as virtual particles in the 
loop (Fig.\ref{fig1}). We will adopt below a terminology where, for example, the 
"gluino-squark contribution" means 
 the sum of those triangle
diagrams, which have a gluino and a squark as virtual states and where a photon or  
a gluon in attached in all possible ways to  the external line. 

Before moving  to derive the two-loop gluino
corrections, we first present, in order to familiarize our notation, the one-loop
results. Those expressions can be found in the literature,
and we just translate them here into our notations for later
convenience. 

The corrections from the gluino-squark diagrams are
\begin{eqnarray}
&&d_{_{\tilde{g}(1)}}^\gamma=-{2\over3\pi}e_{_q}e\alpha_{_s}\sum\limits
_{i=1}^2{\bf Im}\Big(({\cal Z}_{_{\tilde q}})_{_{2,i}}
({\cal Z}_{_{\tilde q}}^\dagger)_{_{i,1}}e^{-i\theta_{_3}}\Big)
\nonumber\\
&&\hspace{1.2cm}\times
{|m_{_{\tilde g}}|\over m_{_{\tilde{q}_i}}^2}B\Big({|m_{_{\tilde g}}|^2
\over m_{_{\tilde{q}_i}}^2}\Big)
\;,\nonumber\\
&&d_{_{\tilde{g}(1)}}^g={g_3\alpha_{_s}\over4\pi}\sum\limits
_{i=1}^2{\bf Im}\Big(({\cal Z}_{_{\tilde q}})_{_{2,i}} ({\cal
Z}_{_{\tilde q}}^\dagger)_{_{i,1}}e^{-i\theta_{_3}}\Big)
\nonumber\\
&&\hspace{1.2cm}\times
{|m_{_{\tilde g}}|\over m_{_{\tilde{q}_i}}^2}C\Big({|m_{_{\tilde g}}|^2
\over m_{_{\tilde{q}_i}}^2}\Big)\;.
\label{eqa3-1}
\end{eqnarray}
Here $\alpha_{_s}=g_3^2/(4\pi)$, $\theta_{_3}$ denotes the phase of 
the soft gluino mass $m_{_{\tilde g}}$, and ${\cal
Z}_{_{\tilde q}}$ are the mixing matrices of the squarks, i.e.
${\cal Z}_{_{\tilde q}}^\dagger{\bf m}_{_{\tilde q}}^2{\cal
Z}_{_{\tilde q}} =diag(m_{_{{\tilde q}_1}}^2,\;m_{_{{\tilde
q}_2}}^2)$ where
\begin{widetext}
\begin{eqnarray}
&&{\bf m}_{_{\tilde q}}^2=\left(\begin{array}{cc}
m_{_{\tilde Q}}^2+m_{_q}^2+m_{_{\rm z}}^2({1\over2}-Q_{_q}s_{_{\rm w}}^2)
\cos2\beta&m_{_q}(A_{_q}^*-\mu R_{_q})\\m_{_q}(A_{_q}-\mu^*R_{_q})&
m_{_{\{{\tilde U},{\tilde D}\}}}^2+m_{_q}^2+m_{_{\rm z}}^2({1\over2}-Q_{_q}s_{_{\rm w}}^2)
\cos2\beta\end{array}\right)\;,
\label{eqa3-a}
\end{eqnarray}
\end{widetext}
with $Q_{_{q}}=2/3(-1/3),\;R_{_q}=\tan\beta(1/\tan\beta)$ for
$q=u\;(d)$. As usual, $\tan\beta={\upsilon_{_u}/\upsilon_{_d}}$ is the
ratio between the VEVs of the up- and down-type Higgs fields, and $\theta_{_{\rm
W}}$ is the weak mixing angle. We also use the short-hand notations
$s_{_{\rm W}}=\sin\theta_{_{\rm W}},\;c_{_{\rm
W}}=\cos\theta_{_{\rm W}}$. The loop functions $B(r)$ and $C(r)$ are defined as
$B(r)=[2(r-1)^2]^{-1} [1+r+2r\ln
r/(r-1)]$ and $C(r)=[6(r-1)^2]^{-1}[10r-26-(2r-18)\ln r/(r-1)]$.

In a similar way, the one-loop neutralino-squark contributions
can be written as
\begin{eqnarray}
&&d_{_{\chi_{_k}^0(1)}}^\gamma=e_{_q}{e\alpha\over16\pi s_{_{\rm w}}^2
c_{_{\rm w}}^2}\sum\limits_{i,k}{\bf Im}\Big((A_{_N}^q)_{_{k,i}}
(B_{_N}^q)^\dagger_{_{i,k}}\Big)
\nonumber\\
&&\hspace{1.2cm}\times
{m_{_{\chi_{_0}^k}}\over m_{_{\tilde{q}_i}}^2}B\Big({m_{_{\chi_{_0}^k}}^2
\over m_{_{\tilde{q}_i}}^2}\Big)
\;,\nonumber\\
&&d_{_{\chi_{_k}^0(1)}}^g={g_3\alpha_{_s}\over64\pi s_{_{\rm w}}^2
c_{_{\rm w}}^2}\sum\limits_{i,k}{\bf Im}\Big((A_{_N}^q)_{_{k,i}}
(B_{_N}^q)^\dagger_{_{i,k}}\Big)
\nonumber\\
&&\hspace{1.2cm}\times
{m_{_{\chi_{_0}^k}}\over m_{_{\tilde{q}_i}}^2}B\Big({m_{_{\chi_{_0}^k}}^2
\over m_{_{\tilde{q}_i}}^2}\Big)
\label{eqa3-2}
\end{eqnarray}
with
\begin{eqnarray}
&&(A_{_N}^u)_{_{k,i}}=-{4\over3}s_{_{\rm w}}({\cal Z}_{_{\tilde u}})_{_{2,i}}
({\cal Z}_{_N})_{_{1,k}}+{m_{_u}c_{_{\rm w}}\over m_{_{\rm w}}s_{_\beta}}
\nonumber\\
&&\hspace{1.8cm}\times
({\cal Z}_{_{\tilde u}})_{_{1,i}}({\cal Z}_{_N})_{_{4,k}}
\;,\nonumber\\
&&(B_{_N}^u)_{_{k,i}}=({\cal Z}_{_{\tilde u}})_{_{1,i}}\Big({s_{_{\rm w}}\over3}
({\cal Z}_{_N})_{_{1,k}}^*+c_{_{\rm w}}({\cal Z}_{_N})_{_{2,k}}^*\Big)
\nonumber\\
&&\hspace{1.8cm}
+{m_{_u}c_{_{\rm w}}\over m_{_{\rm w}}s_{_\beta}}
({\cal Z}_{_{\tilde u}})_{_{2,i}}({\cal Z}_{_N})_{_{4,k}}^*
\;,\nonumber\\
&&(A_{_N}^d)_{_{k,i}}={2\over3}s_{_{\rm w}}({\cal Z}_{_{\tilde d}})_{_{2,i}}
({\cal Z}_{_N})_{_{1,k}}+{m_{_d}c_{_{\rm w}}\over m_{_{\rm w}}c_{_\beta}}
\nonumber\\
&&\hspace{1.8cm}\times
({\cal Z}_{_{\tilde d}})_{_{1,i}}({\cal Z}_{_N})_{_{3,k}}
\;,\nonumber\\
&&(B_{_N}^d)_{_{k,i}}=({\cal Z}_{_{\tilde d}})_{_{1,i}}\Big({s_{_{\rm w}}\over3}
({\cal Z}_{_N})_{_{1,k}}^*-c_{_{\rm w}}({\cal Z}_{_N})_{_{2,k}}^*\Big)
\nonumber\\
&&\hspace{1.8cm}
+{m_{_d}c_{_{\rm w}}\over m_{_{\rm w}}c_{_\beta}}
({\cal Z}_{_{\tilde d}})_{_{2,i}}({\cal Z}_{_N})_{_{3,k}}^*\;.
\label{eqa3-3}
\end{eqnarray}
Here $\alpha=e^2/(4\pi),\;s_{_\beta}=\sin\beta,\;c_{_\beta}=\cos\beta$, 
$m_{_{\chi_{_0}^k}}\;(k=1,\;2,\;3,\;4)$ denote the eigenvalues of 
neutralino mass matrix, and ${\cal Z}_{_N}$ is the correspondingly 
mixing matrix.

\begin{figure}[b]
\setlength{\unitlength}{1mm}
\begin{center}
\begin{picture}(0,80)(0,0)
\put(-50,-50){\includegraphics{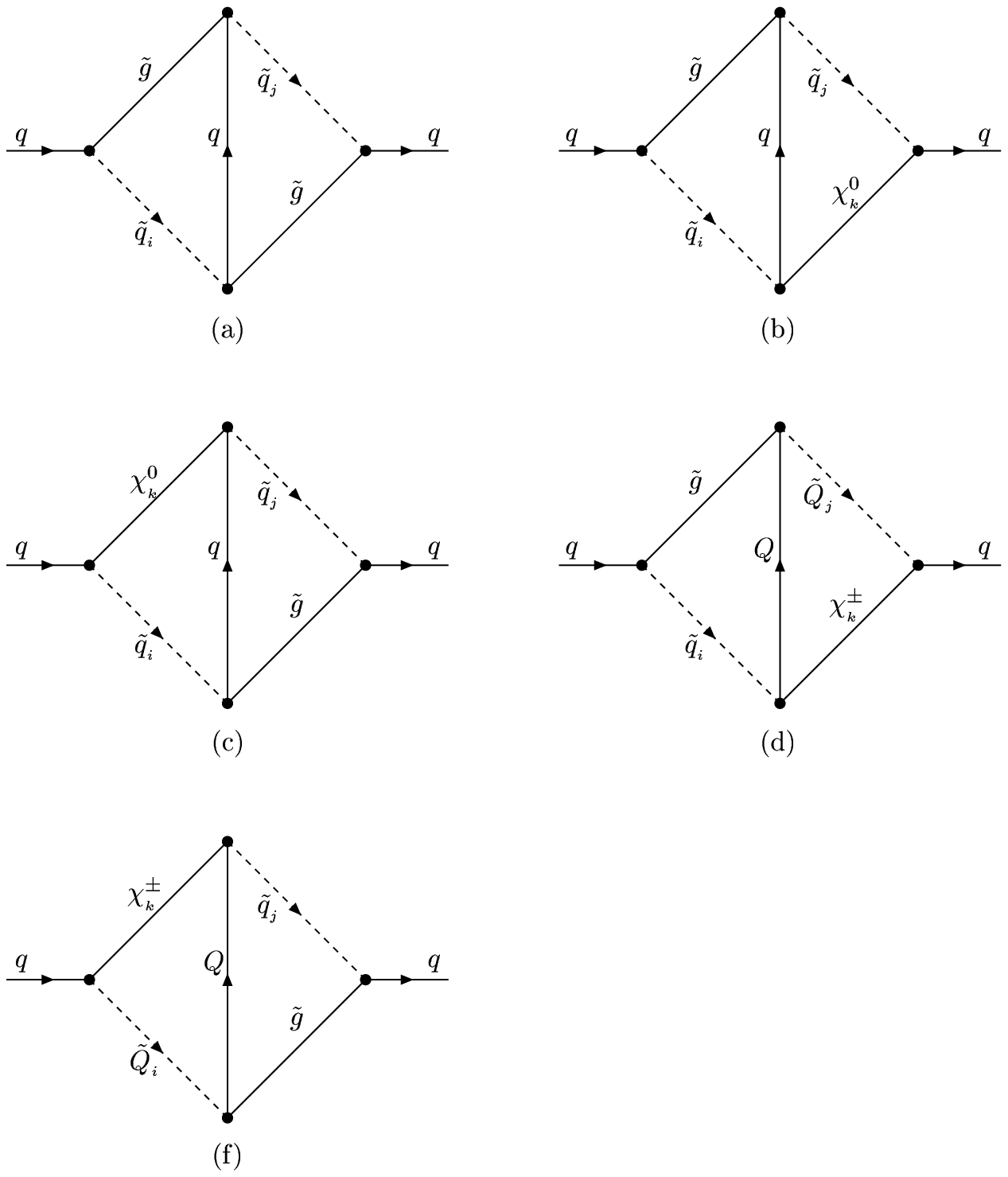}}
\end{picture}
\caption[]{The two-loop self energy diagrams which lead to the
quark EDMs and CEDMs in the MSSM, the corresponding triangle
diagrams are obtained by attaching a photon or  gluon in all
possible ways to the SUSY particles.} \label{fig2}
\end{center}
\end{figure}

Finally, the chargino-squark contributions are given by
\begin{eqnarray}
&&d_{_{\chi_{_k}^\pm(1)}}^\gamma={e\alpha\over4\pi s_{_{\rm w}}^2}
V_{_{qQ}}^\dagger V_{_{Qq}}
\sum\limits_{i,k}{\bf Im}\Big((A_{_C}^Q)_{_{k,i}}(B_{_C}^Q)^\dagger
_{_{i,k}}\Big){m_{_{\chi_{_k}^\pm}}\over m_{_{\tilde{Q}_i}}^2}
\nonumber\\
&&\hspace{1.2cm}\times
\Big[e_{_Q}B\Big({m_{_{\chi_{_k}^\pm}}^2\over m_{_{\tilde{Q}_i}}^2}\Big)
+(e_{_q}-e_{_Q})A\Big({m_{_{\chi_{_k}^\pm}}^2\over
m_{_{\tilde{Q}_i}}^2}\Big)\Big]
\;,\nonumber\\
&&d_{_{\chi_{_k}^\pm(1)}}^g={g_3\alpha\over4\pi s_{_{\rm w}}^2}
V_{_{qQ}}^\dagger V_{_{Qq}}
\sum\limits_{i,k}{\bf Im}\Big((A_{_C}^Q)_{_{k,i}}(B_{_C}^Q)^\dagger
_{_{i,k}}\Big)
\nonumber\\
&&\hspace{1.2cm}\times
{m_{_{\chi_{_k}^\pm}}\over m_{_{\tilde{Q}_i}}^2}
B\Big({m_{_{\chi_{_k}^\pm}}^2\over m_{_{\tilde{Q}_i}}^2}\Big)\;,
\label{eqa3-4}
\end{eqnarray}
where $V$ denotes the CKM matrix, $m_{_{\chi_{_k}^\pm}}\;(k=1,\;2)$
are the masses of charginos, and the loop function
$A(r)=2(1-r)^{-2}[3-r+2\ln r/(1-r)]$.
The couplings appearing in these expressions are defined as
\begin{eqnarray}
&&(A_{_C}^d)_{_{k,i}}={m_{_u}\over\sqrt{2}m_{_{\rm w}}s_{_\beta}}
({\cal Z}_{_{\tilde d}})_{_{1,i}}({\cal Z}_+)_{_{2,k}}
\;,\nonumber\\
&&(B_{_C}^d)_{_{k,i}}={m_{_d}\over\sqrt{2}m_{_{\rm w}}c_{_\beta}}
({\cal Z}_{_{\tilde d}})_{_{2,i}}({\cal Z}_-)_{_{2,k}}
-({\cal Z}_{_{\tilde d}})_{_{1,i}}({\cal Z}_-)_{_{1,k}}
\;,\nonumber\\
&&(A_{_C}^u)_{_{k,i}}={m_{_d}\over\sqrt{2}m_{_{\rm w}}c_{_\beta}}
({\cal Z}_{_{\tilde u}})_{_{1,i}}({\cal Z}_-)_{_{2,k}}^*
\;,\nonumber\\
&&(B_{_C}^u)_{_{k,i}}={m_{_u}\over\sqrt{2}m_{_{\rm w}}s_{_\beta}}
({\cal Z}_{_{\tilde u}})_{_{2,i}}({\cal Z}_+)_{_{2,k}}^*
-({\cal Z}_{_{\tilde u}})_{_{1,i}}({\cal Z}_+)_{_{1,k}}^*
\;,\nonumber\\
\label{eqa3-5}
\end{eqnarray}
where ${\cal Z}_+$ and ${\cal Z}_-$ are the right- and left-handed mixing
matrices of the charginos, respectively. It is noteworthy that one-loop chargino
corrections to the quark EDMs and CEDMs are proportional to the
suppression factors ${m_{_q}/m_{_{\rm w}}}$ contained in the
couplings $A_{_C}^q \;(q=u,\;d)$.

The two-loop gluino corrections to the Wilson coefficients originate
from the two-loop self-energy diagrams for quarks, depicted in Fig. \ref{fig2}. 
The corresponding dipole moment diagrams are obtained by attaching a photon
or gluon to these diagrams in all possible ways. In these diagrams there is no new
suppression factor, except  a factor arising from loop integration, 
and the divergence caused by the sub-diagrams is subtracted in 
$\overline{MS}$ scheme safely. It turns out that for some region 
of the parameter space the two-loop results are comparable with 
the one-loop contributions. The reason for this is
that the dependence of the two-loop results on the relevant
CP phases differs from that of the one-loop results. 

Since the two-loop analysis is more subtle than the analysis at one-loop level, we
present here in some detail all the processes, which  contribute at two-loop
level into the theoretical prediction of the quark EDMs. Taking
the same steps, which we did in our earlier works \cite{step}, we
obtain the following expression for the relevant effective Lagrangian:
\begin{eqnarray}
&&{\cal L}_{_{eff}}^{\tilde{g}\tilde{g}}=
{4\over9}g_3^4\int{d^Dq_1\over(2\pi)^D}{d^Dq_2\over(2\pi)^D}
{1\over{\cal D}_{_{\tilde g}}}
\nonumber\\
&&\hspace{1.2cm}\times
\Big\{({\cal Z}_{_{\tilde q}})_{_{2,j}}
({\cal Z}_{_{\tilde q}}^\dagger)_{_{j,1}}({\cal Z}_{_{\tilde q}})_{_{1,i}}
({\cal Z}_{_{\tilde q}}^\dagger)_{_{i,1}}|m_{_{\tilde g}}|e^{-i\theta_{_3}}
\Big[e_{_q}e{\cal N}_{_{\tilde{g}(1)}}^\gamma{\cal Q}_{_1}
+g_3{\cal N}_{_{\tilde{g}(1)}}^g{\cal Q}_{_3}\Big]
\nonumber\\
&&\hspace{1.2cm}
+({\cal Z}_{_{\tilde q}})_{_{1,j}}
({\cal Z}_{_{\tilde q}}^\dagger)_{_{j,2}}({\cal Z}_{_{\tilde q}})_{_{2,i}}
({\cal Z}_{_{\tilde q}}^\dagger)_{_{i,2}}|m_{_{\tilde g}}|e^{i\theta_{_3}}
\Big[e_{_q}e{\cal N}_{_{\tilde{g}(1)}}^\gamma{\cal Q}_{_2}
+g_3{\cal N}_{_{\tilde{g}(1)}}^g{\cal Q}_{_4}\Big]
\nonumber\\
&&\hspace{1.2cm}
+({\cal Z}_{_{\tilde q}})_{_{2,j}}
({\cal Z}_{_{\tilde q}}^\dagger)_{_{j,2}}({\cal Z}_{_{\tilde q}})_{_{2,i}}
({\cal Z}_{_{\tilde q}}^\dagger)_{_{i,1}}|m_{_{\tilde g}}|e^{-i\theta_{_3}}
\Big[e_{_q}e{\cal N}_{_{\tilde{g}(2)}}^\gamma{\cal Q}_{_1}
+g_3{\cal N}_{_{\tilde{g}(2)}}^g{\cal Q}_{_3}\Big]
\nonumber\\
&&\hspace{1.2cm}
+({\cal Z}_{_{\tilde q}})_{_{1,j}}
({\cal Z}_{_{\tilde q}}^\dagger)_{_{j,1}}({\cal Z}_{_{\tilde q}})_{_{1,i}}
({\cal Z}_{_{\tilde q}}^\dagger)_{_{i,2}}|m_{_{\tilde g}}|e^{i\theta_{_3}}
\Big[e_{_q}e{\cal N}_{_{\tilde{g}(2)}}^\gamma{\cal Q}_{_2}
+g_3{\cal N}_{_{\tilde{g}(2)}}^g{\cal Q}_{_4}\Big]\Big\}+\cdots\;,
\label{eq4}
\end{eqnarray}
where ${\cal D}_{_{\tilde
g}}=((q_2-q_1)^2-m_{_q}^2)(q_1^2-|m_{_{\tilde g}}|^2)
(q_1^2-m_{_{\tilde{q}_j}}^2)(q_2^2-m_{_{\tilde{q}_i}}^2)
(q_2^2-|m_{_{\tilde g}}|^2)$ and $D$ is the time-space dimension. We
collect the complicated form factor ${\cal N}_{_{\tilde{g}}}$ in
appendix \ref{ap1}. 

The effective Lagrangian of Eq. (\ref{eq4})
originates in fact from the contributions of the effective triangle
vertices induced by the "gluino-gluino self-energy" (Fig.2
(a)). The terminology here is analogous to the one adopted in the one-loop case 
above. The
gluino-gluino self-energy is a short-hand expression for a diagram, where
we attach a photon or gluon line in all possible ways to the
two-loop quark self-energy diagram having, besides the squark,
two gluinos are as intermediate agents. Because the sum of the triangle diagrams
corresponding to each "self-energy" obviously respects the gauge
invariance, we can calculate the contributions of all the
"self-energies" separately. We will use below the following identities
demanded by the translational invariance of the loop-momenta:
\begin{eqnarray}
&&\int{d^Dq_1\over(2\pi)^D}{d^Dq_2\over(2\pi)^D}{1\over{\cal D}_{_0}}
\Big\{-{2+D\over D}q_1\cdot(q_2-q_1)
\nonumber\\&&
+{2\over q_2^2-m_{_2}^2}\Big[{D(q_1\cdot q_2)^2-q_1^2q_2^2\over D(D-1)}
-{q_1^2q_1\cdot q_2\over D}\Big]
\nonumber\\&&
+{2\over q_1^2-m_{_1}^2}
{q_1^2q_1\cdot(q_2-q_1)\over D}
\Big\}\equiv0\;,
\nonumber\\
&&\int{d^Dq_1\over(2\pi)^D}{d^Dq_2\over(2\pi)^D}{1\over{\cal D}_{_0}}
\Big\{-q_1\cdot(q_2-q_1)
\nonumber\\&&
+{2\over q_1^2-m_{_1}^2}\Big[{D(q_1\cdot q_2)^2-q_1^2q_2^2\over D(D-1)}
-{q_1^2q_1\cdot q_2\over D}\Big]
\nonumber\\&&
+{2\over q_2^2-m_{_2}^2}{q_1\cdot(q_2-q_1)q_2^2\over D}
\Big\}\equiv0\;,
\label{eq5}
\end{eqnarray}
where ${\cal D}_{_0}=((q_2-q_1)^2-m_{_0}^2)(q_1^2-m_{_1}^2)(q_2^2-m_{_2}^2)$.
We find
\begin{widetext}
\begin{eqnarray}
&&\int{d^Dq_1\over(2\pi)^D}{d^Dq_2\over(2\pi)^D}
{{\cal N}_{_{\tilde{g}(2)}}^\gamma\over{\cal D}_{_{\tilde g}}}
\Big(\tilde{q}_i\leftrightarrow \tilde{q}_j,\;q_1\leftrightarrow q_2\Big)
\nonumber\\
=&&\hspace{0.0cm}
\int{d^Dq_1\over(2\pi)^D}{d^Dq_2\over(2\pi)^D}{1\over{\cal D}_{_{\tilde g}}}
\Big\{{2\over D}
{q_1\cdot(q_1-q_2)\over q_1^2-m_{_{\tilde{q}_j}}^2}+{4\over D}
{q_2^2q_1\cdot(q_1-q_2)\over(q_2^2-m_{_{\tilde{q}_i}}^2)^2}
-{q_1\cdot(q_1-q_2)\over q_2^2-m_{_{\tilde{q}_i}}^2}
\nonumber\\
&&\hspace{0.0cm}
+{4\over(q_2^2-m_{_{\tilde{q}_i}}^2)(q_1^2-|m_{_{\tilde g}}|^2)}
\Big[{q_1\cdot q_2q_1^2\over D}
-{D(q_1\cdot q_2)^2-q_1^2q_2^2\over D(D-1)}\Big]
+{4\over D}{q_1^2q_1\cdot(q_1-q_2)\over(q_1^2-|m_{_{\tilde g}}|^2)^2}
\nonumber\\
&&\hspace{0.0cm}
-{2+D\over D}{q_1\cdot(q_1-q_2)\over q_1^2-|m_{_{\tilde g}}|^2}
+{2-D\over D}{q_1\cdot(q_1-q_2)\over(q_2-q_1)^2-m_{_q}^2}\Big\}
\nonumber\\
=&&\hspace{.0cm}
\int{d^Dq_1\over(2\pi)^D}{d^Dq_2\over(2\pi)^D}{1\over{\cal D}_{_{\tilde g}}}
\Big\{{D-2\over D}{q_1\cdot(q_2-q_1)\over(q_2-q_1)^2-m_{_q}^2}
+{2\over D}{q_1^2q_1\cdot(q_2-q_1)\over(q_1^2-|m_{_{\tilde g}}|^2)
(q_1^2-m_{_{\tilde{q}_j}}^2)}
\nonumber\\
&&\hspace{0.0cm}
+\Big[{2\over(q_1^2-|m_{_{\tilde g}}|^2)
(q_2^2-|m_{_{\tilde g}}|^2)}+{1\over(q_1^2-m_{_{\tilde{q}_j}}^2)
(q_2^2-m_{_{\tilde{q}_i}}^2)}\Big]\Big[
{D(q_1\cdot q_2)^2-q_1^2q_2^2\over D(D-1)}
\nonumber\\
&&\hspace{.0cm}
-{q_1^2q_1\cdot q_2\over D}\Big]
+{2\over D}{q_2^2q_1\cdot(q_2-q_1)\over(q_2^2-m_{_{\tilde{q}_i}}^2)
(q_2^2-|m_{_{\tilde g}}|^2)}
-{2\over D}{q_1\cdot(q_2-q_1)\over q_1^2-m_{_{\tilde{q}_j}}^2}\Big\}
\nonumber\\
=&&\hspace{.0cm}
\int{d^Dq_1\over(2\pi)^D}{d^Dq_2\over(2\pi)^D}{1\over{\cal D}_{_{\tilde g}}}
\Big\{{D-2\over D}{q_1\cdot(q_2-q_1)\over(q_2-q_1)^2-m_{_q}^2}
-{4\over D}{q_1\cdot(q_2-q_1)q_2^2\over(q_2^2-|m_{_{\tilde g}}|^2)^2}
\nonumber\\
&&\hspace{.0cm}
-{4\over(q_1^2-m_{_{\tilde{q}_j}}^2)(q_2^2-|m_{_{\tilde g}}|^2)}\Big[
{D(q_1\cdot q_2)^2-q_1^2q_2^2\over D(D-1)}
-{q_1^2q_1\cdot q_2\over D}\Big]
\nonumber\\
&&\hspace{.0cm}
-{4\over D}{q_1^2q_1\cdot(q_2-q_1)\over(q_1^2-m_{_{\tilde{q}_j}}^2)^2}
+{q_1\cdot(q_2-q_1)\over q_2^2-|m_{_{\tilde g}}|^2}
+{q_1\cdot(q_2-q_1)\over q_1^2-m_{_{\tilde{q}_j}}^2}\Big\}
\nonumber\\
=&&\hspace{.0cm}
\int{d^Dq_1\over(2\pi)^D}{d^Dq_2\over(2\pi)^D}
{{\cal N}_{_{\tilde{g}(1)}}^\gamma\over{\cal D}_{_{\tilde g}}}\;.
\label{eq6}
\end{eqnarray}
\end{widetext}
Similarly, we find
\begin{eqnarray}
&&\int{d^Dq_1\over(2\pi)^D}{d^Dq_2\over(2\pi)^D}
{{\cal N}_{_{\tilde{g}(2)}}^g\over{\cal D}_{_{\tilde g}}}
\Big(\tilde{q}_i\leftrightarrow \tilde{q}_j,\;q_1\leftrightarrow q_2\Big)
\nonumber\\
=&&\hspace{.0cm}
\int{d^Dq_1\over(2\pi)^D}{d^Dq_2\over(2\pi)^D}
{{\cal N}_{_{\tilde{g}(1)}}^g\over{\cal D}_{_{\tilde g}}}\;.
\label{eq7}
\end{eqnarray}
With the help of Eq.(\ref{eq6}) and Eq.(\ref{eq7}), the corresponding terms
in Eq.(\ref{eq4}) are transformed into
\begin{eqnarray}
&&{\cal L}_{_{eff}}^{\tilde{g}\tilde{g}}=-i{4\over9}e_{_q}eg_3^4|m_{_{\tilde g}}|
\int{d^Dq_1\over(2\pi)^D}{d^Dq_2\over(2\pi)^D}
{{\cal N}_{_{\tilde{g}(1)}}^\gamma\over{\cal D}_{_{\tilde g}}}
{\bf Im}\Big(({\cal Z}_{_{\tilde q}})_{_{2,j}}({\cal Z}_{_{\tilde q}}^\dagger)_{_{j,1}}
e^{-i\theta_{_3}}\Big)\cdot\Big[\bar{q}\sigma^{\mu\nu}\gamma_5qF_{\mu\nu}\Big]
\nonumber\\
&&\hspace{1.2cm}
-i{4\over9}g_3^5|m_{_{\tilde g}}|\int{d^Dq_1\over(2\pi)^D}
{d^Dq_2\over(2\pi)^D}{{\cal N}_{_{\tilde{g}(1)}}^g\over{\cal D}_{_{\tilde g}}}
{\bf Im}\Big(({\cal Z}_{_{\tilde q}})_{_{2,j}}({\cal Z}_{_{\tilde q}}^\dagger)_{_{j,1}}
e^{-i\theta_{_3}}\Big)\cdot\Big[\bar{q}T^a\sigma^{\mu\nu}\gamma_5qG^a_{\mu\nu}\Big]
\nonumber\\
&&\hspace{1.2cm}
+\cdots\;.
\label{eq7a}
\end{eqnarray}
The two-loop scalar integrations can be reduced into the two-loop
vacuum integrals  defined as
$$\int{d^Dq_1d^Dq_2\over(q_1-q_2)^2-m_0^2)(q_1^2-m_1^2)(q_2^2-m_2^2)},$$
which was analyzed in detail \cite{2vac}.  
However, the two-loop scalar integration $\int{d^Dq_1\over(2\pi)^D}
{d^Dq_2\over(2\pi)^D}{{\cal N}_{_{\tilde{g}(1)}}^{\gamma,g}\over{\cal D}_{_{\tilde g}}}$
contain the ultra-violet divergence which originates from the sub-daigrams.
After the renormalization in $\overline{MS}$ scheme, we finally obtain the
corrections from the two-loop "gluino-gluino" diagram  to the
quark EDMs and CEDMs respectively as
\begin{eqnarray}
&&d_{_{\tilde{g}(2)}}^\gamma={8e_{_q}e\alpha_{_s}^2|m_{_{\tilde g}}|\over9(4\pi)^2m_{_{\rm w}}^2}
F_1(x_{_q}, x_{_{\tilde{q}_j}},x_{_{\tilde g}},x_{_{\tilde g}},x_{_{\tilde{q}_i}})
{\bf Im}\Big(({\cal Z}_{_{\tilde q}})_{_{2,j}}({\cal Z}_{_{\tilde q}}^\dagger)_{_{j,1}}
e^{-i\theta_{_3}}\Big)
\;,\nonumber\\
&&d_{_{\tilde{g}(2)}}^g={8g_3\alpha_{_s}^2|m_{_{\tilde g}}|\over9(4\pi)^2m_{_{\rm w}}^2}
F_3(x_{_q}, x_{_{\tilde{q}_j}},x_{_{\tilde g}},x_{_{\tilde g}},x_{_{\tilde{q}_i}})
{\bf Im}\Big(({\cal Z}_{_{\tilde q}})_{_{2,j}}({\cal Z}_{_{\tilde q}}^\dagger)_{_{j,1}}
e^{-i\theta_{_3}}\Big)\;,
\label{eq8}
\end{eqnarray}
with $x_i=m_i^2/m_{_{\rm w}}^2$, and the function $F_i(x_0,x_1
,x_2,x_3,x_4)\;(i=1,\;3)$ are defined in appendix \ref{ap2}.

The effective
Lagrangian, which corresponds to the contributions of the triangle
diagrams induced in the "neutralino-gluino" self-energy
(i.e. diagrams where loop squarks are accompanied with neutralinos and gluinos, 
see Fig.2 (b) and (c) ), includes the following pieces:
\begin{eqnarray}
&&{\cal L}_{_{eff}}^{\tilde{g}\chi_{_k}^0}=
{2\over3}{e^2g_3^2\over s_{_{\rm w}}^2c_{_{\rm w}}^2}\int
{d^Dq_1\over(2\pi)^D}{d^Dq_2\over(2\pi)^D}{1\over{\cal D}_{_{\chi_{_k}^0}}^{(a)}}
\nonumber\\
&&\hspace{1.2cm}\times
\Big\{(A_{_N}^q)_{_{kj}}({\cal Z}_{_{\tilde q}}^\dagger)_{_{j,1}}
(B_{_N}^q)_{_{ki}}({\cal Z}_{_{\tilde q}}^\dagger)_{_{i,1}}|m_{_{\tilde g}}|
e^{-i\theta_{_3}}\Big[e_{_q}e{\cal N}_{_{\tilde{g}(1)}}^\gamma{\cal Q}_{_1}
+g_3{\cal N}_{_{\chi_{_k}^0(1)}}^{g(a)}{\cal Q}_{_3}\Big]
\nonumber\\
&&\hspace{1.2cm}
+(B_{_N}^q)_{_{kj}}({\cal Z}_{_{\tilde q}}^\dagger)_{_{j,2}}
(A_{_N}^q)_{_{ki}}({\cal Z}_{_{\tilde q}}^\dagger)_{_{i,2}}|m_{_{\tilde g}}|
e^{i\theta_{_3}}\Big[e_{_q}e{\cal N}_{_{\tilde{g}(1)}}^\gamma{\cal Q}_{_2}
+g_3{\cal N}_{_{\chi_{_k}^0(1)}}^{g(a)}{\cal Q}_{_4}\Big]
\nonumber\\
&&\hspace{1.2cm}
-(A_{_N}^q)_{_{kj}}({\cal Z}_{_{\tilde q}}^\dagger)_{_{j,2}}(A_{_N}^q)_{_{ki}}
({\cal Z}_{_{\tilde q}}^\dagger)_{_{i,1}}m_{_{\chi_{_k}^0}}
\Big[e_{_q}e{\cal N}_{_{\tilde{g}(2)}}^\gamma{\cal Q}_{_1}
+g_3{\cal N}_{_{\chi_{_k}^0(2)}}^{g(a)}{\cal Q}_{_3}\Big]
\nonumber\\
&&\hspace{1.2cm}
-(B_{_N}^q)_{_{kj}}({\cal Z}_{_{\tilde q}}^\dagger)_{_{j,1}}(B_{_N}^q)_{_{ki}}
({\cal Z}_{_{\tilde q}}^\dagger)_{_{i,2}}m_{_{\chi_{_k}^0}}
\Big[e_{_q}e{\cal N}_{_{\tilde{g}(2)}}^\gamma{\cal Q}_{_2}
+g_3{\cal N}_{_{\chi_{_k}^0(2)}}^{g(a)}{\cal Q}_{_4}\Big]\Big\}
\nonumber\\
&&\hspace{1.2cm}
+{2\over3}{e^2g_3^2\over s_{_{\rm w}}^2c_{_{\rm w}}^2}\int
{d^Dq_1\over(2\pi)^D}{d^Dq_2\over(2\pi)^D}{1\over{\cal D}_{_{\chi_{_k}^0}}^{(b)}}
\nonumber\\
&&\hspace{1.2cm}\times
\Big\{-({\cal Z}_{_{\tilde q}})_{_{2,j}}(B_{_N}^q)_{_{jk}}^\dagger
({\cal Z}_{_{\tilde q}})_{_{1,i}}(B_{_N}^q)_{_{ik}}^\dagger m_{_{\chi_{_k}^0}}
\Big[e_{_q}e{\cal N}_{_{\chi_{_k}^0(1)}}^{\gamma(b)}{\cal Q}_{_1}
+g_3{\cal N}_{_{\chi_{_k}^0(1)}}^{g(b)}{\cal Q}_{_3}\Big]
\nonumber\\
&&\hspace{1.2cm}
-({\cal Z}_{_{\tilde q}})_{_{1,j}}(A_{_N}^q)_{_{jk}}^\dagger
({\cal Z}_{_{\tilde q}})_{_{2,i}}(A_{_N}^q)_{_{ik}}^\dagger m_{_{\chi_{_k}^0}}
\Big[e_{_q}e{\cal N}_{_{\chi_{_k}^0(1)}}^{\gamma(b)}{\cal Q}_{_2}
+g_3{\cal N}_{_{\chi_{_k}^0(1)}}^{g(b)}{\cal Q}_{_4}\Big]
\nonumber\\
&&\hspace{1.2cm}
+({\cal Z}_{_{\tilde q}})_{_{2,j}}(A_{_N}^q)_{_{jk}}^\dagger
({\cal Z}_{_{\tilde q}})_{_{2,i}}(B_{_N}^q)_{_{ik}}^\dagger
|m_{_{\tilde g}}|e^{-i\theta_{_3}}
\Big[e_{_q}e{\cal N}_{_{\chi_{_k}^0(2)}}^{\gamma(b)}{\cal Q}_{_1}
+g_3{\cal N}_{_{\chi_{_k}^0(2)}}^{g(b)}{\cal Q}_{_3}\Big]
\nonumber\\
&&\hspace{1.2cm}
+({\cal Z}_{_{\tilde q}})_{_{1,j}}(B_{_N}^q)_{_{jk}}^\dagger
({\cal Z}_{_{\tilde q}})_{_{1,i}}(A_{_N}^q)_{_{ik}}^\dagger
|m_{_{\tilde g}}|e^{i\theta_{_3}}
\Big[e_{_q}e{\cal N}_{_{\chi_{_k}^0(2)}}^{\gamma(b)}{\cal Q}_{_2}
+g_3{\cal N}_{_{\chi_{_k}^0(2)}}^{g(b)}{\cal Q}_{_4}\Big]\Big\}+\cdots
\label{eq9}
\end{eqnarray}
with ${\cal
D}_{_{\chi_{_k}^0}}^{(a)}=((q_2-q_1)^2-m_{_q}^2)(q_1^2-m_{_{\chi_{_k}^0}}^2)
(q_1^2-m_{_{\tilde{q}_j}}^2)(q_2^2-m_{_{\tilde{q}_i}}^2)(q_2^2-|m_{_{\tilde g}}|^2),\;
{\cal D}_{_{\chi_{_k}^0}}^{(b)}={\cal
D}_{_{\chi_{_k}^0}}^{(a)}(m_{_{\chi_{_k}^0}}^2 \leftrightarrow
|m_{_{\tilde g}}|^2)$. The form factors ${\cal
N}_{_{\chi_{_k}^0(1,2)}}^{\gamma(b)} ={\cal
N}_{_{\tilde{g}(1,2)}}^\gamma(|m_{_{\tilde g}}|^2\rightarrow
m_{_{\chi_{_k}^0}}^2)$. The explicit expressions of the form
factors ${\cal N}_{_{\chi_{_k}^0(1,2)}}^{g(a,b)}$ are presented in
appendix \ref{ap1}.

With a help of Eq. (\ref{eq5}), one finds the following identities:
\begin{eqnarray}
&&\int{d^Dq_1\over(2\pi)^D}{d^Dq_2\over(2\pi)^D}{{\cal N}_{_{\chi_{_k}^0(1)}}^{\gamma(b)}
,{\cal N}_{_{\chi_{_k}^0(1)}}^{g(b)}\over{\cal D}_{_{\chi_{_k}^0}}^{(b)}}
\Big(\tilde{q}_i\leftrightarrow \tilde{q}_j,\;q_1\leftrightarrow q_2\Big)
\nonumber\\
=&&\hspace{.0cm}
\int{d^Dq_1\over(2\pi)^D}{d^Dq_2\over(2\pi)^D}{{\cal N}_{_{\tilde{g}(2)}}^\gamma,
{\cal N}_{_{\chi_{_k}^0(2)}}^{g(a)}\over{\cal D}_{_{\chi_{_k}^0}}^{(a)}}
\;,\nonumber\\
&&\int{d^Dq_1\over(2\pi)^D}{d^Dq_2\over(2\pi)^D}{{\cal N}_{_{\chi_{_k}^0(1)}}^{\gamma(b)},
{\cal N}_{_{\chi_{_k}^0(1)}}^{g(b)}\over{\cal D}_{_{\chi_{_k}^0}}^{(b)}}
\Big(\tilde{q}_i\leftrightarrow \tilde{q}_j,\;q_1\leftrightarrow q_2\Big)
\nonumber\\
=&&\hspace{.0cm}
\int{d^Dq_1\over(2\pi)^D}{d^Dq_2\over(2\pi)^D}{{\cal N}_{_{\tilde{g}(2)}}^\gamma,
{\cal N}_{_{\chi_{_k}^0(2)}}^{g(a)}\over{\cal D}_{_{\chi_{_k}^0}}^{(a)}}
\;,\nonumber\\
&&\int{d^Dq_1\over(2\pi)^D}{d^Dq_2\over(2\pi)^D}{{\cal N}_{_{\chi_{_k}^0(2)}}^{\gamma(b)},
{\cal N}_{_{\chi_{_k}^0(2)}}^{g(b)}\over{\cal D}_{_{\chi_{_k}^0}}^{(b)}}
\Big(\tilde{q}_i\leftrightarrow \tilde{q}_j,\;q_1\leftrightarrow q_2\Big)
\nonumber\\
=&&\hspace{.0cm}
\int{d^Dq_1\over(2\pi)^D}{d^Dq_2\over(2\pi)^D}{{\cal N}_{_{\tilde{g}(1)}}^\gamma,
{\cal N}_{_{\chi_{_k}^0(1)}}^{g(a)}\over{\cal D}_{_{\chi_{_k}^0}}^{(a)}}
\;,\nonumber\\
&&\int{d^Dq_1\over(2\pi)^D}{d^Dq_2\over(2\pi)^D}{{\cal N}_{_{\chi_{_k}^0(2)}}^{\gamma(b)},
{\cal N}_{_{\chi_{_k}^0(2)}}^{g(b)}\over{\cal D}_{_{\chi_{_k}^0}}^{(b)}}
\Big(\tilde{q}_i\leftrightarrow \tilde{q}_j,\;q_1\leftrightarrow q_2\Big)
\nonumber\\
=&&\hspace{.0cm}
\int{d^Dq_1\over(2\pi)^D}{d^Dq_2\over(2\pi)^D}{{\cal N}_{_{\tilde{g}(1)}}^\gamma,
{\cal N}_{_{\chi_{_k}^0(1)}}^{g(a)}\over{\cal D}_{_{\chi_{_k}^0}}^{(a)}}\;.
\label{eq10}
\end{eqnarray}
Substituting these identities into eq.(15) and removing the ultra-violet
divergence in $\overline{MS}$ scheme, we may extract  the two-loop
"neutralino-gluino" corrections to the quark EDMs and CEDMs:
\begin{eqnarray}
&&d_{_{\chi_{_k}^0(2)}}^\gamma=
{4e_{_q}e\alpha\alpha_{_s}\over3(4\pi)^2s_{_{\rm w}}^2
c_{_{\rm w}}^2m_{_{\rm w}}^2}\Big\{|m_{_{\tilde g}}|F_1(x_{_q}, 
x_{_{\tilde{q}_j}}, x_{_{\tilde g}},x_{_{\chi_k^0}}, 
x_{_{\tilde{q}_i}})\Big[{\bf Im}\Big((A_{_N}^q)_{_{kj}}
({\cal Z}_{_{\tilde q}}^\dagger)_{_{j,1}}
\nonumber\\
&&\hspace{1.4cm}\times
(B_{_N}^q)_{_{ki}}({\cal Z}_{_{\tilde q}}^\dagger)_{_{i,1}}
e^{-i\theta_{_3}}\Big)-{\bf Im}\Big((B_{_N}^q)_{_{kj}}
({\cal Z}_{_{\tilde q}}^\dagger)_{_{j,2}}(A_{_N}^q)_{_{ki}}
({\cal Z}_{_{\tilde q}}^\dagger)_{_{i,2}}e^{i\theta_{_3}}\Big)\Big]
\nonumber\\
&&\hspace{1.4cm}
-m_{_{\chi_{_k}^0}}F_2(x_{_q}, x_{_{\tilde{q}_j}}, x_{_{\tilde g}},
x_{_{\chi_k^0}}, x_{_{\tilde{q}_i}})\Big[{\bf Im}
\Big((A_{_N}^q)_{_{kj}}({\cal Z}_{_{\tilde q}}^\dagger)_{_{j,2}}
(A_{_N}^q)_{_{ki}}({\cal Z}_{_{\tilde q}}^\dagger)_{_{i,1}}\Big)
\nonumber\\
&&\hspace{1.4cm}
-{\bf Im}\Big((B_{_N}^q)_{_{kj}}({\cal Z}_{_{\tilde q}}^\dagger)_{_{j,1}}
(B_{_N}^q)_{_{ki}}({\cal Z}_{_{\tilde q}}^\dagger)_{_{i,2}}\Big)\Big]\Big\}
\;,\nonumber\\
&&d_{_{\chi_{_k}^0(2)}}^g=
{4g_3\alpha\alpha_{_s}\over3(4\pi)^2s_{_{\rm w}}^2c_{_{\rm w}}^2m_{_{\rm w}}^2}
\Big\{|m_{_{\tilde g}}|F_4(x_{_q}, x_{_{\tilde{q}_j}}, x_{_{\tilde g}},
x_{_{\chi_k^0}}, x_{_{\tilde{q}_i}})\Big[{\bf Im}\Big((A_{_N}^q)_{_{kj}}
({\cal Z}_{_{\tilde q}}^\dagger)_{_{j,1}}
\nonumber\\
&&\hspace{1.4cm}\times
(B_{_N}^q)_{_{ki}}({\cal Z}_{_{\tilde q}}^\dagger)_{_{i,1}}
e^{-i\theta_{_3}}\Big)-{\bf Im}\Big((B_{_N}^q)_{_{kj}}
({\cal Z}_{_{\tilde q}}^\dagger)_{_{j,2}}(A_{_N}^q)_{_{ki}}
({\cal Z}_{_{\tilde q}}^\dagger)_{_{i,2}}e^{i\theta_{_3}}\Big)\Big]
\nonumber\\
&&\hspace{1.4cm}
-m_{_{\chi_{_k}^0}}F_5(x_{_q}, x_{_{\tilde{q}_j}}, x_{_{\tilde g}},
x_{_{\chi_k^0}}, x_{_{\tilde{q}_i}})\Big[{\bf Im}
\Big((A_{_N}^q)_{_{kj}}({\cal Z}_{_{\tilde q}}^\dagger)_{_{j,2}}
(A_{_N}^q)_{_{ki}}({\cal Z}_{_{\tilde q}}^\dagger)_{_{i,1}}\Big)
\nonumber\\
&&\hspace{1.4cm}
-{\bf Im}\Big((B_{_N}^q)_{_{kj}}({\cal Z}_{_{\tilde q}}^\dagger)_{_{j,1}}
(B_{_N}^q)_{_{ki}}({\cal Z}_{_{\tilde q}}^\dagger)_{_{i,2}}\Big)\Big]
\Big\}\;.
\label{eq11}
\end{eqnarray}

In a similar way, we get the following results for the two-loop "gluino-chargino"
corrections:
\begin{eqnarray}
&&d_{_{\chi_{_k}^\pm(2)}}^\gamma=
{4e\alpha\alpha_{_s}\over3(4\pi)^2s_{_{\rm w}}^2m_{_{\rm w}}^2}V_{_{qQ}}^\dagger V_{_{Qq}}
\Big\{|m_{_{\tilde g}}|F_6(x_{_Q}, x_{_{\tilde{Q}_j}}, x_{_{\tilde g}},
x_{_{\chi_k^\pm}}, x_{_{\tilde{q}_i}})\Big[{\bf Im}\Big((A_{_C}^Q)_{_{kj}}
({\cal Z}_{_{\tilde Q}}^\dagger)_{_{j,1}}
\nonumber\\
&&\hspace{1.4cm}\times
(B_{_C}^q)_{_{ki}}({\cal Z}_{_{\tilde q}}^\dagger)_{_{i,1}}
e^{-i\theta_{_3}}\Big)-{\bf Im}\Big((B_{_C}^Q)_{_{kj}}
({\cal Z}_{_{\tilde Q}}^\dagger)_{_{j,2}}
(A_{_C}^q)_{_{ki}}({\cal Z}_{_{\tilde q}}^\dagger)_{_{i,2}}
e^{i\theta_{_3}}\Big)\Big]
\nonumber\\
&&\hspace{1.4cm}
-m_{_{\chi_{_k}^\pm}}F_7(x_{_Q}, x_{_{\tilde{Q}_j}}, x_{_{\tilde g}},
x_{_{\chi_k^\pm}}, x_{_{\tilde{q}_i}})\Big[{\bf Im}
\Big((A_{_C}^Q)_{_{kj}}({\cal Z}_{_{\tilde Q}}^\dagger)_{_{j,2}}
(A_{_C}^q)_{_{ki}}({\cal Z}_{_{\tilde q}}^\dagger)_{_{i,1}}\Big)
\nonumber\\
&&\hspace{1.4cm}
-{\bf Im}\Big((B_{_C}^Q)_{_{kj}}({\cal Z}_{_{\tilde Q}}^\dagger)_{_{j,1}}
(B_{_C}^q)_{_{ki}}({\cal Z}_{_{\tilde q}}^\dagger)_{_{i,2}}\Big)\Big]\Big\}
\;,\nonumber\\
&&d_{_{\chi_{_k}^\pm(2)}}^g=
{4g_3\alpha\alpha_{_s}\over3(4\pi)^2s_{_{\rm w}}^2m_{_{\rm w}}^2}V_{_{qQ}}^\dagger V_{_{Qq}}
\Big\{|m_{_{\tilde g}}|F_4(x_{_Q}, x_{_{\tilde{Q}_j}}, x_{_{\tilde g}},
x_{_{\chi_k^\pm}}, x_{_{\tilde{q}_i}})\Big[{\bf Im}\Big((A_{_C}^Q)_{_{kj}}
({\cal Z}_{_{\tilde Q}}^\dagger)_{_{j,1}}
\nonumber\\
&&\hspace{1.4cm}\times
(B_{_C}^q)_{_{ki}}({\cal Z}_{_{\tilde q}}^\dagger)_{_{i,1}}
e^{-i\theta_{_3}}\Big)
-{\bf Im}\Big((B_{_C}^Q)_{_{kj}}({\cal Z}_{_{\tilde Q}}^\dagger)_{_{j,2}}
(A_{_C}^q)_{_{ki}}({\cal Z}_{_{\tilde q}}^\dagger)_{_{i,2}}
e^{i\theta_{_3}}\Big)\Big]
\nonumber\\
&&\hspace{1.4cm}
-m_{_{\chi_{_k}^\pm}}F_5(x_{_Q}, x_{_{\tilde{Q}_j}}, x_{_{\tilde g}},
x_{_{\chi_k^\pm}}, x_{_{\tilde{q}_i}})\Big[{\bf Im}
\Big((A_{_C}^Q)_{_{kj}}({\cal Z}_{_{\tilde Q}}^\dagger)_{_{j,2}}
(A_{_C}^q)_{_{ki}}({\cal Z}_{_{\tilde q}}^\dagger)_{_{i,1}}\Big)
\nonumber\\
&&\hspace{1.4cm}
-{\bf Im}\Big((B_{_C}^Q)_{_{kj}}({\cal Z}_{_{\tilde Q}}^\dagger)_{_{j,1}}
(B_{_C}^q)_{_{ki}}({\cal Z}_{_{\tilde q}}^\dagger)_{_{i,2}}\Big)\Big]
\Big\}\;.\nonumber\\
\label{eq12}
\end{eqnarray}
Notice that the last terms of the $d_{_{\chi_{_k}^\pm(2)}}^\gamma$
and $d_{_{\chi_{_k}^\pm(2)}}^g$ are not proportional to the
suppression factor $m_{_q}/m_{_{\rm w}}$. This implies that the
two-loop "gluino-chargino" diagrams may be  dominant two-loop
corrections of the chargino to the quark EDMs and CEDMs.

The Wilson coefficient of the purely gluonic Weinberg operator
originates from the two-loop "gluino-squark" diagrams  
and is given by \cite{Dai}:
\begin{eqnarray}
&&C_{_5}=-3\alpha_{_s}m_{_t}\Big({g_3\over4\pi}\Big)^3
{\bf Im}\Big(({\cal Z}_{_{\tilde t}})_{_{2,2}}
({\cal Z}_{_{\tilde t}})^\dagger_{_{2,1}}\Big)
\nonumber\\
&&\hspace{1.2cm}\times
{m_{_{\tilde{t}_1}}^2-m_{_{\tilde{t}_2}}^2\over |m_{_{\tilde g}}|^5}
H({m_{_{\tilde{t}_1}}^2\over |m_{_{\tilde g}}|^2},{m_{_{\tilde{t}_2}}^2
\over |m_{_{\tilde g}}|^2},{m_{_t}^2\over |m_{_{\tilde g}}|^2})
\label{eq13}
\end{eqnarray}
where ${\cal Z}_{_{\tilde t}}$ is the diagonalizing matrix for the
squared mass matrix of stop, and the function $H$ is the same as
that given in \cite{Dai}.

In order to account for the resummation of logarithmic
corrections, we should evolve the quark EDMs, CEDMs and the Wilson
coefficients of the Weinberg operator with the renormalization group equations 
(RGEs) from the matching scale $\Lambda$
down to the chirality breaking scale $\Lambda_{_\chi}$ \cite{rge}:
\begin{eqnarray}
&&d_{_q}^\gamma(\Lambda_{_\chi})=\eta_{_\gamma}d_{_q}^\gamma(\Lambda)
\;,\nonumber\\
&&d_{_q}^g(\Lambda_{_\chi})=\eta_{_g}d_{_q}^g(\Lambda)
\;,\nonumber\\
&&C_{_5}(\Lambda_{_\chi})=\eta_{_G}C_{_5}(\Lambda)\;,
\label{eq14}
\end{eqnarray}
where $\eta_{_\gamma}\simeq1.53$ and $\eta_{_g}\simeq\eta_{_G}\simeq3.4$. 
At a low scale, the quark EDM consists of a weak interaction contribution, a quark
CEDMs contribution and the contribution of the Weinberg operator. We can 
numerically  evaluate these different contributions. According to a naive 
dimensional analysis one can write \cite{nda} 
\begin{eqnarray}
&&d_{_q}=d_{_q}^\gamma+{e\over4\pi}d_{_q}^g+{e\Lambda_{_\chi}
\over4\pi}C_{_5}(\Lambda_{_\chi})\;.
\label{eq15}
\end{eqnarray}
The EDM of the neutron is, on the basis of the simple SU(6) quark model, then 
given by
\begin{eqnarray}
&&d_{_n}={1\over3}(4d_{_d}-d_{_u})\;,
\label{eq16}
\end{eqnarray}
where $d_{_u},\;d_{_d}$ are the EDMs of the up- and down-type
quarks respectively.

\section{The numerical analysis \label{sec3}}
\indent\indent The MSSM Lagrangian contains several  sources for  CP violating 
phases: the phases of the $\mu$
parameter in the superpotential and the corresponding bilinear
coupling of the soft breaking terms, three phases of the gaugino
mass terms, and the phases of the trilinear sfermion Yukawa couplings in
the soft Lagrangian. As we are not considering the spontaneous CP
violation in this work, the CP phase of soft bilinear coupling
vanishes due to the tree level neutral Higgs tadpole conditions.

As we remarked above, the two-loop gluino corrections to the
neutron EDM are important, and  may be the dominant contribution
in a certain part of the parameter space.
Let us clarify this point in some detail by means of 
numerical analysis. Without losing too much generality, we will fix
the following values for the supersymmetric parameters:
$|\mu|=|m_{_{\lambda_B}}|=|m_{_{\lambda_A}}|=|m_{_{\tilde g}}|=300\;{\rm GeV}$,
\footnote{$m_{_{\lambda_B}},\;m_{_{\lambda_A}}$ are the soft masses of $U(1)
\times SU(2)$ gauginos respectively.}
$|A_{_q}|=100\;{\rm GeV} \;(q=u,\;d,\;c,\;s,\;t,\;b)$, and
$m_{_{\tilde{Q}_2}}=20\;{\rm TeV},\; m_{_{\tilde{Q}_3}}=5\;{\rm
TeV},\;m_{_{\tilde C}}=m_{_{\tilde S}} =5\;{\rm TeV},\;m_{_{\tilde
T}}=m_{_{\tilde B}}=200\;{\rm GeV}$ in our numerical analysis. For
simplification, we will also assume 
$\varphi_{_q}=arg(A_{_q})=0,\;\theta_{_\mu}=arg(\mu)=0$.

\begin{figure}
\setlength{\unitlength}{1mm}
\begin{center}
\begin{picture}(0,80)(0,0)
\put(-50,-30){\includegraphics{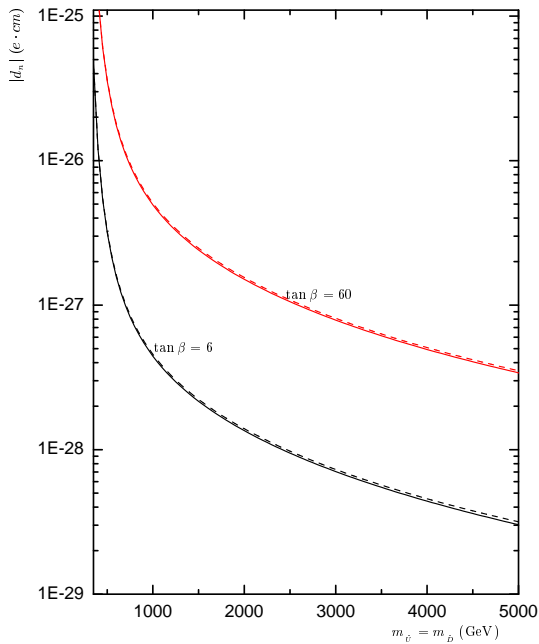}}
\end{picture}
\caption[]{The correction from one-loop gluino diagrams (Eq.
\ref{eqa3-1}) and that from two-loop gluino-gluino diagrams (Eq.
\ref{eq8}) versus the right handed squark masses $m_{_{\tilde
U}}=m_{_{\tilde D}}$ with
$\theta_{_3}=arg(m_{_{\tilde g}})=\pi/2,\;\tan\beta=6\;(60)$, and
$m_{_{\tilde{Q}_1}}=20\;{\rm TeV}$, where the dash-lines represent
one-loop results  and the solid-lines correspond to a sum of the
one- and two-loop results.} \label{fig3}
\end{center}
\end{figure}

In Fig. \ref{fig3} we present the corrections
the neutron EDM obtains  from the one-loop "gluino-squark" diagrams  (see Eq.
(\ref{eqa3-1}) and Fig.~1) and
two-loop "gluino-gluino" contributions (see Eq. (\ref{eq8}) and Fig.~2) as a 
function of the
right-handed squark masses $m_{_{\tilde U}}=m_{_{\tilde D}}$. We have set in this 
figure $\theta_{_1}=arg(m_{_{\lambda_B}})=0,\;\theta_{_2}=
arg(m_{_{\lambda_A}})=0,\;\theta_{_3}=arg(m_{_{\tilde g}})=
\pi/2,\;\tan\beta=6\;(60)$, and
$m_{_{\tilde{Q}_1}}=20\;{\rm TeV}$, and what are plotted are the one-loop 
"gluino-squark" contribution (dashed line) and the  sum of the one-loop 
"gluino-squark" and two-loop "gluino-gluino" contributions (solid line). The 
dependence of the contributions on
the CP violation phase $\theta_{_3}$  are all proportional to
the factor $ {\bf Im} \Big(({\cal Z}_{_{\tilde q}})_{_{2,j}}({\cal
Z}_{_{\tilde q}}^\dagger)_{_{j,1}} e^{-i\theta_{_3}}\Big)$, and
the two-loop correction is less than 10\% of the one-loop results
for our choice of the parameter values.  The neutron EDM depends on the parameter
$\tan\beta$ through the squark mixing
matrices, and as seen from the plots, this makes the a large difference between 
the corrections in a low and a high $\tan\beta$ cases.

\begin{figure}
\setlength{\unitlength}{1mm}
\begin{center}
\begin{picture}(0,80)(0,0)
\put(-50,-30){\includegraphics{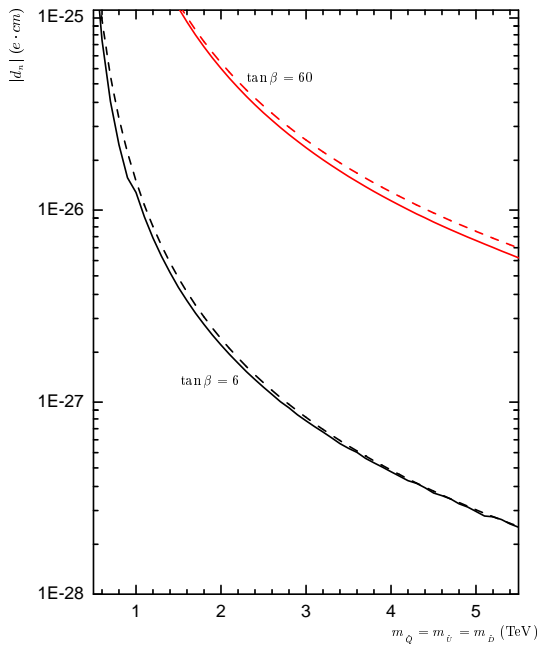}}
\end{picture}
\caption[]{The correction from one-loop neutralino diagrams (Eq.
\ref{eqa3-2}) and that from two-loop neutralino-gluino diagrams
(Eq. \ref{eq11}) versus the squark masses
$m_{_{\tilde{Q}_1}}=m_{_{\tilde U}}=m_{_{\tilde D}}$ with
$\theta_{_1}=\pi/2,\;\theta_{_2}=\theta_{_3}=0,\;
\tan\beta=6\;(60)$, where the dash-lines represent one-loop
results  and the solid-lines correspond to a sum of the one- and
two-loop results.} \label{fig4}
\end{center}
\end{figure}

Let us now focus on the two-loop gluino corrections of the
neutron EDM. There the CP violating phases
$\theta_{_1},\;\theta_{_2}$ induce a nonzero quark EDMs at
one-loop level (see Eq. (\ref{eqa3-2})) through the neutralino mixing
matrix, whereas the CP phase $\theta_{_3}$ contributes only
through the two-loop diagrams (see Eq. (\ref{eq11})). We will present three sets 
of plots, where we set  one of the phases $\theta_1,\theta_2,\theta_3$ at a time 
equal to $\pi/2$ while the other two are set equal to zero. Taking first
$\theta_{_1}=\pi/2,\; \theta_{_2}=\theta_{_3}=0$ and
$\tan\beta=6\;(60)$, we plot in Fig. \ref{fig4} for two values of $\tan\beta$  the 
neutron EDM as a function of the 
soft squark masses $m_{_{\tilde{Q}_1}}=m_{_{\tilde U}}=m_{_{\tilde D}}$.
In this plot the dash-lines represent the one-loop
"neutralino-squark" results and the solid-lines correspond to a
sum of one-loop "neutralino-squark" and two-loop
"neutralino-gluino" contributions. 

\begin{figure}
\setlength{\unitlength}{1mm}
\begin{center}
\begin{picture}(0,80)(0,0)
\put(-50,-30){\includegraphics{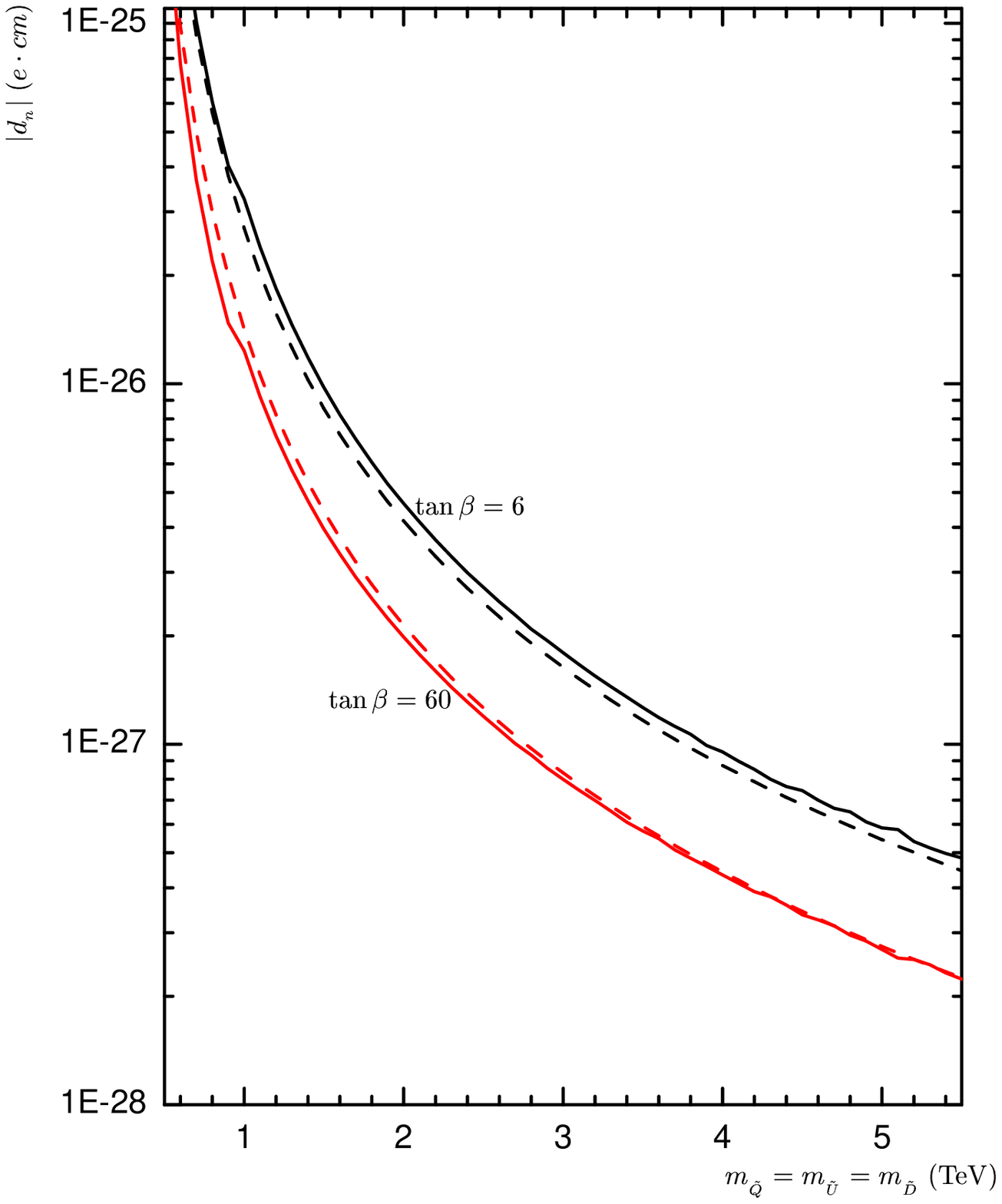}}
\end{picture}
\caption[]{The correction from one-loop neutralino diagrams (Eq.
\ref{eqa3-2}) only and the sum of one-loop neutralino and two-loop
neutralino-gluino diagrams (Eq. \ref{eq11}) versus the squark
masses $m_{_{\tilde{Q}_1}}=m_{_{\tilde U}}=m_{_{\tilde D}}$ with
$\theta_{_2}=\pi/2,\;\theta_{_1}=\theta_{_3}=0,\;
\tan\beta=6\;(60)$, where the dash-lines represent one-loop
results  and the solid-lines correspond to a sum of the one- and
two-loop results.} \label{fig5}
\end{center}
\end{figure}

In Fig. \ref{fig5} we present the same plot for the case
$\theta_{_2}=\pi/2,\; \theta_{_1}=\theta_{_3}=0$, other parameters
being the same as in Fig. \ref{fig4}. We notice that the two-loop
gluino corrections to the one-loop results can be as large as
about $20\%$. For the contribution of neutralino  to the neutron
EDM, the most interesting piece originates from the CP violating
phase of the gluino mass, which does not contribute at the
one-loop level.
 
\begin{figure}
\setlength{\unitlength}{1mm}
\begin{center}
\begin{picture}(0,80)(0,0)
\put(-50,-30){\includegraphics{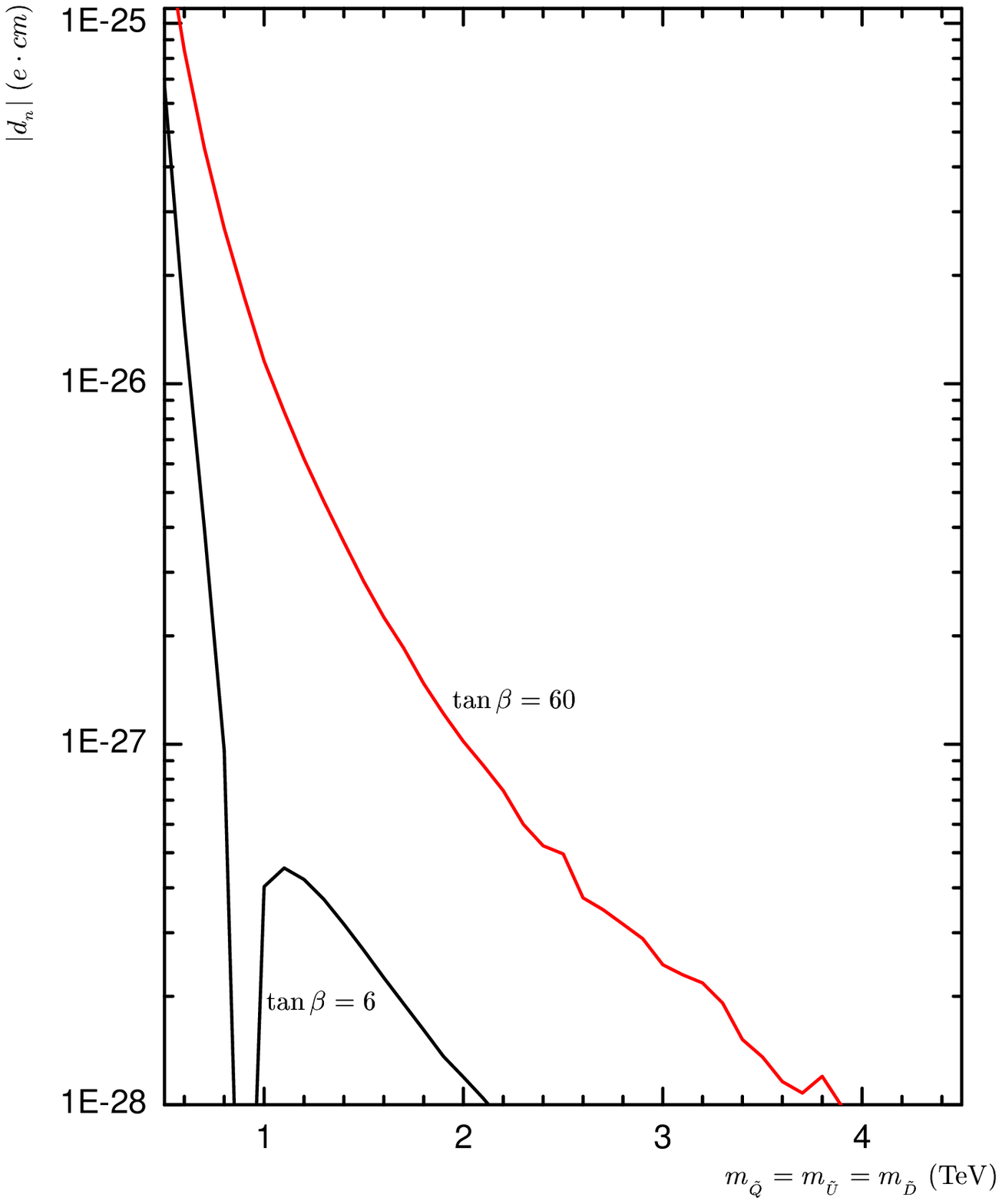}}
\end{picture}
\caption[]{The correction the two-loop
neutralino-gluino diagrams (Eq. \ref{eq11}) versus the
squark masses $m_{_{\tilde{Q}_1}}=m_{_{\tilde U}}=m_{_{\tilde D}}$ with
$\theta_{_3}=\pi/2,\;\theta_{_1}=\theta_{_2}=0,\;
\tan\beta=6\;(60)$. With the choice on parameter space, the one-loop
neutralino contribution to neutron EDM is zero.}
\label{fig6}
\end{center}
\end{figure}

In Fig. \ref{fig6} we plot the neutron EDM versus
the  soft squark masses $m_{_{\tilde{Q}_1}}= m_{_{\tilde
U}}=m_{_{\tilde D}}$ in the case $\theta_{_3}=\pi/2,\;
\theta_{_1}=\theta_{_2}=0$, the other parameters are taken as in
Fig. \ref{fig4}. As the  soft squark masses increase, the neutron
EDM $d_{_n}$ varies steeply. Because of the interference between the
up- and down-quark EDMs (see Eq. \ref{eq16}), there is in the case of  
$\tan\beta=6$ a profound
minimum at $m_{_{\tilde{Q}_1}}= m_{_{\tilde U}}=m_{_{\tilde
D}}\sim 900\;{\rm GeV}$ .

\begin{figure}
\setlength{\unitlength}{1mm}
\begin{center}
\begin{picture}(0,80)(0,0)
\put(-50,-30){\includegraphics{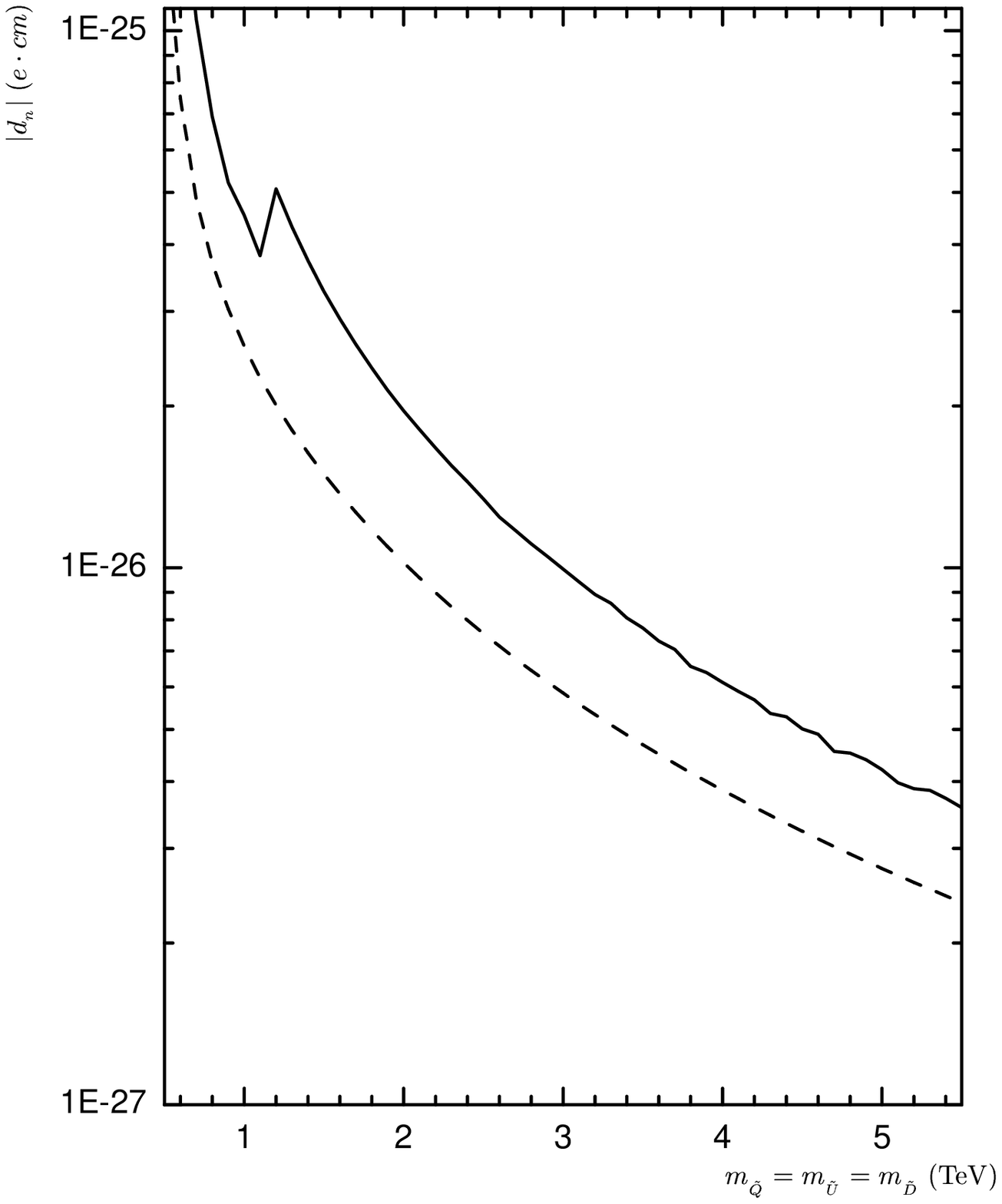}}
\end{picture}
\caption[]{The correction from
one-loop chargino diagrams (Eq. \ref{eqa3-4}) and the sum of one- and
two-loop chargino-gluino diagrams (Eq. \ref{eq12}) versus the
squark masses $m_{_{\tilde{Q}_1}}=m_{_{\tilde U}}=m_{_{\tilde D}}$ with
$\theta_{_2}=\pi/2,\;\theta_{_3}=0,\;\tan\beta=60$.}
\label{fig7}
\end{center}
\end{figure}
\begin{figure}
\setlength{\unitlength}{1mm}
\begin{center}
\begin{picture}(0,80)(0,0)
\put(-50,-30){\includegraphics{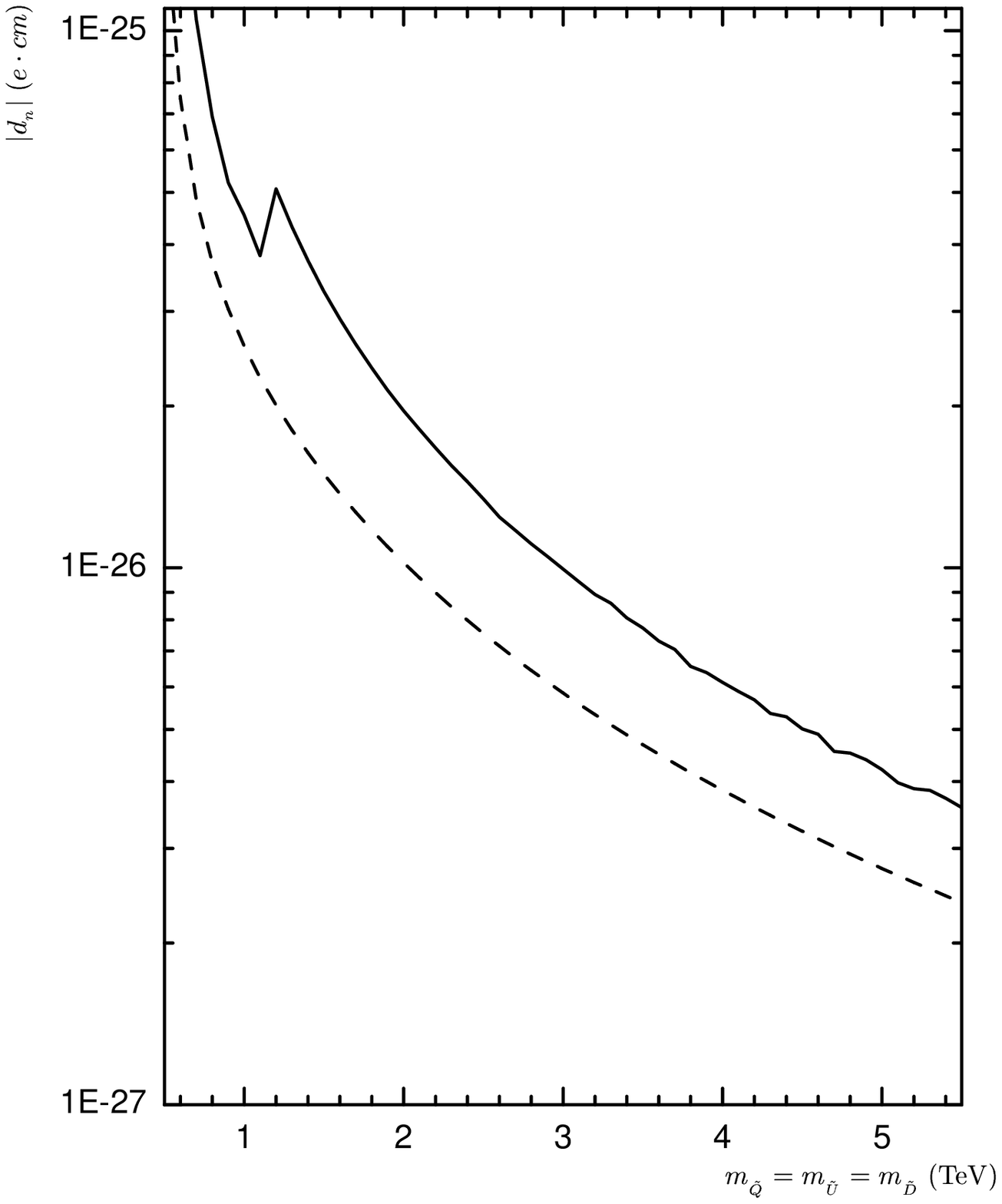}}
\end{picture}
\caption[]{The correction from
one-loop chargino diagrams (Eq. \ref{eqa3-4}) and the sum of one- and
two-loop chargino-gluino diagrams (Eq. \ref{eq12}) versus the
squark masses $m_{_{\tilde{Q}_1}}=m_{_{\tilde U}}=m_{_{\tilde D}}$ with
$\theta_{_2}=\pi/2,\;\theta_{_3}=0,\;\tan\beta=6$.}
\label{fig8}
\end{center}
\end{figure}
\begin{figure}
\setlength{\unitlength}{1mm}
\begin{center}
\begin{picture}(0,80)(0,0)
\put(-50,-30){\includegraphics{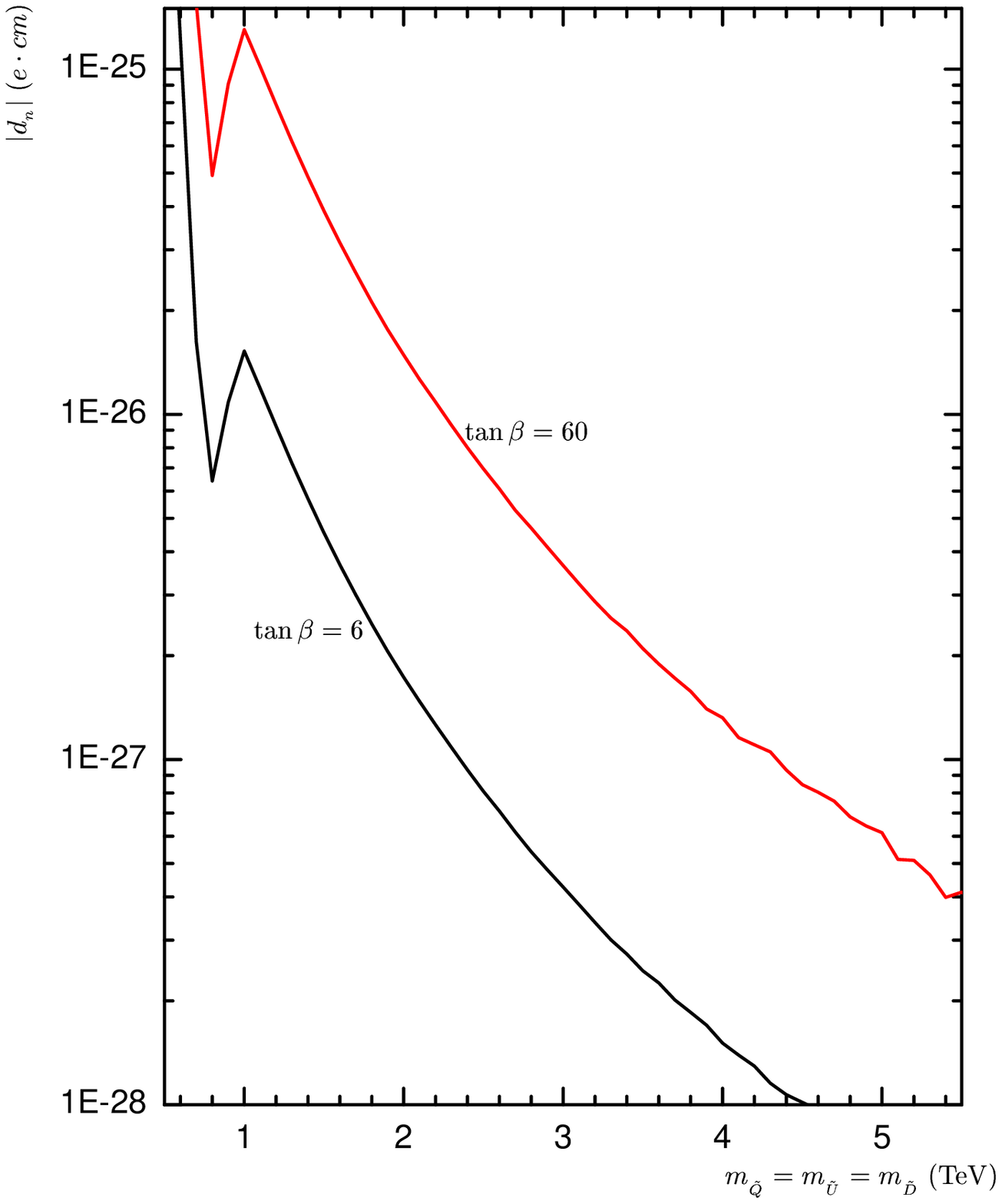}}
\end{picture}
\caption[]{The correction from the
two-loop chargino-gluino diagrams (Eq. \ref{eq12}) versus the
squark masses $m_{_{\tilde{Q}_1}}=m_{_{\tilde U}}=m_{_{\tilde D}}$ with
$\theta_{_3}=\pi/2,\;\theta_{_1}=\theta_{_2}=0,\;
\tan\beta=6, \;(60)$.}
\label{fig9}
\end{center}
\end{figure}
Let us move to the corrections arising from the diagrams involving virtual 
charginos. In this case the phase $\theta_{_1}$ does
not contribute to the quark EDMs at all, whereas $\theta_{_2}$ 
contributes at  one-loop level (see Eq. (\ref{eqa3-4})) and
$\theta_{_3}$ takes part in the game through only two-loop
diagrams (see Eq. (\ref{eq12})). At the one-loop level, the chargino
contributions to the quark EDMs are proportional to the suppression
factor ${m_{_q}} /m_{_{\rm z}}\;(q=u,\;d)$. However, the last
terms of the two-loop "gluino-chargino" corrections to the quark EDMs and CEDMs, 
given in Eq.
(\ref{eq12}), do not depend on this
suppression factor. Taking
$\theta_{_2}=\pi/2,\;\theta_{_3}=0$, we plot in Fig.
\ref{fig7} the
neutron EDM as a function of the soft squark masses
$m_{_{\tilde{Q}_1}}=m_{_{\tilde U}}=m_{_{\tilde D}}$ for a large $\tan\beta$. The 
dash lines represent the one-loop
"chargino-squark" results  and the solid lines correspond to the sum
of the one-loop "chargino-squark" and two-loop "chargino-gluino"
results. As the dominant two-loop "gluino-chargino" corrections
are not proportional to the suppression factor given above, they play the
leading role. The corresponding figure for a low $\tan\beta$ is presented in Fig.
\ref{fig8}. 

Just as in the neutralino case, the
CP phase $\theta_{_3}$ affects also in the chargino case the quark EDMs only at 
the two-loop
level. In Fig. \ref{fig9} we show the variation of the  neutron EDM with
respect to $m_{_{\tilde{Q}_1}}=m_{_{\tilde U}}=m_{_{\tilde D}}$
in the case $\theta_{_3}=\pi/2,\;\theta_{_2}=0$ and $\tan\beta=6\;(60)$.
Even when the squark masses are larger than $1\;{\rm TeV}$, the
neutron EDM induced by the CP phase $\theta_{_3}$ can be close
to $1.1\times10^{-25}\; (e\cdot cm)$, which is the present upper
bound from experiments.
\begin{figure}
\setlength{\unitlength}{1mm}
\begin{center}
\begin{picture}(0,80)(0,0)
\put(-50,-30){\includegraphics{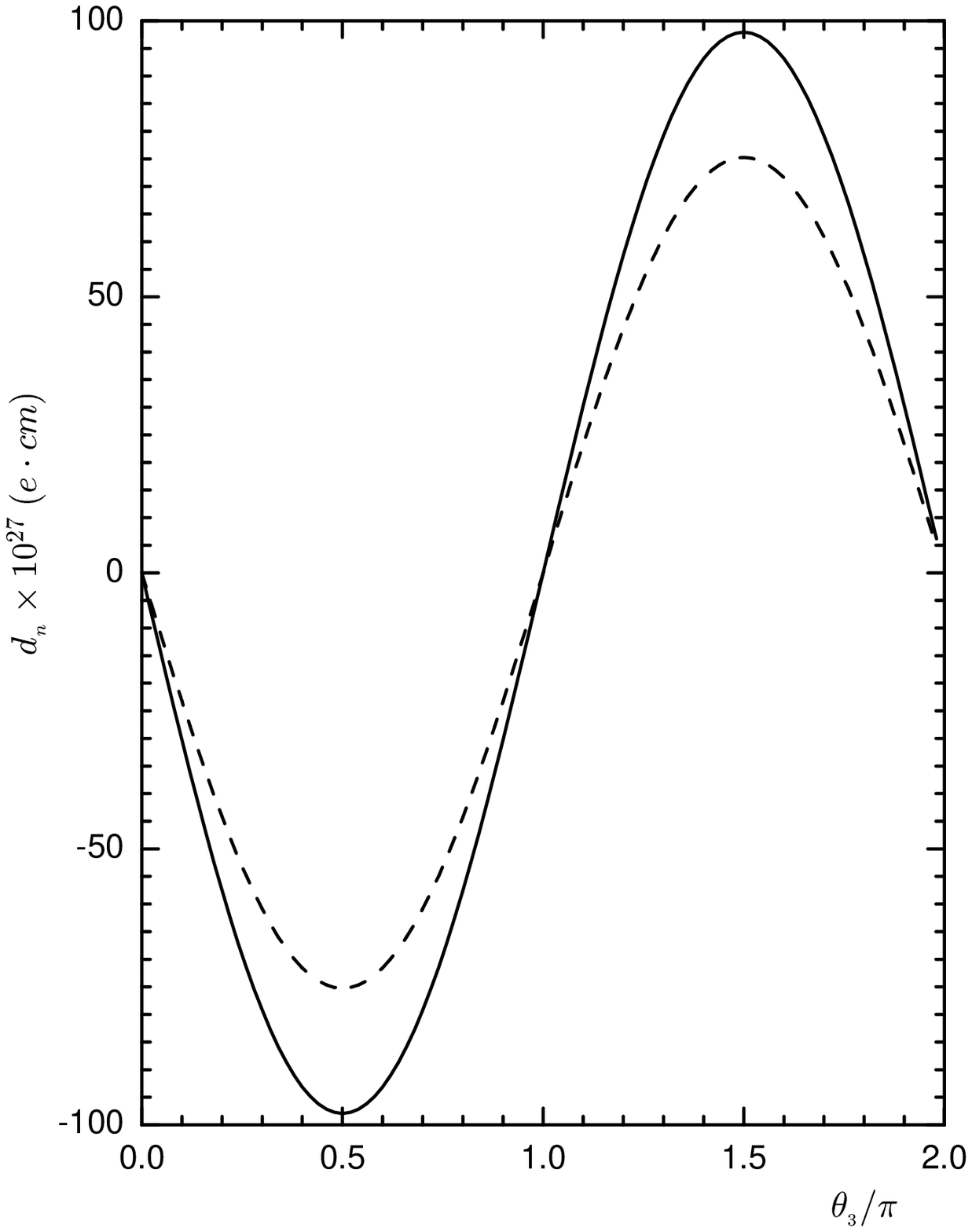}}
\end{picture}
\caption[]{Taking $\theta_{_1}=\theta_{_2}=0,\;m_{_{\tilde{Q}_1}}
=m_{_{\tilde U}}=m_{_{\tilde D}}=2.3\;{\rm TeV}$ and $\tan\beta=60$,
the neutron EDM varies with the $CP$ phase $\theta_{_3}$.}
\label{fig10}
\end{center}
\end{figure}
\begin{figure}
\setlength{\unitlength}{1mm}
\begin{center}
\begin{picture}(0,80)(0,0)
\put(-50,-30){\includegraphics{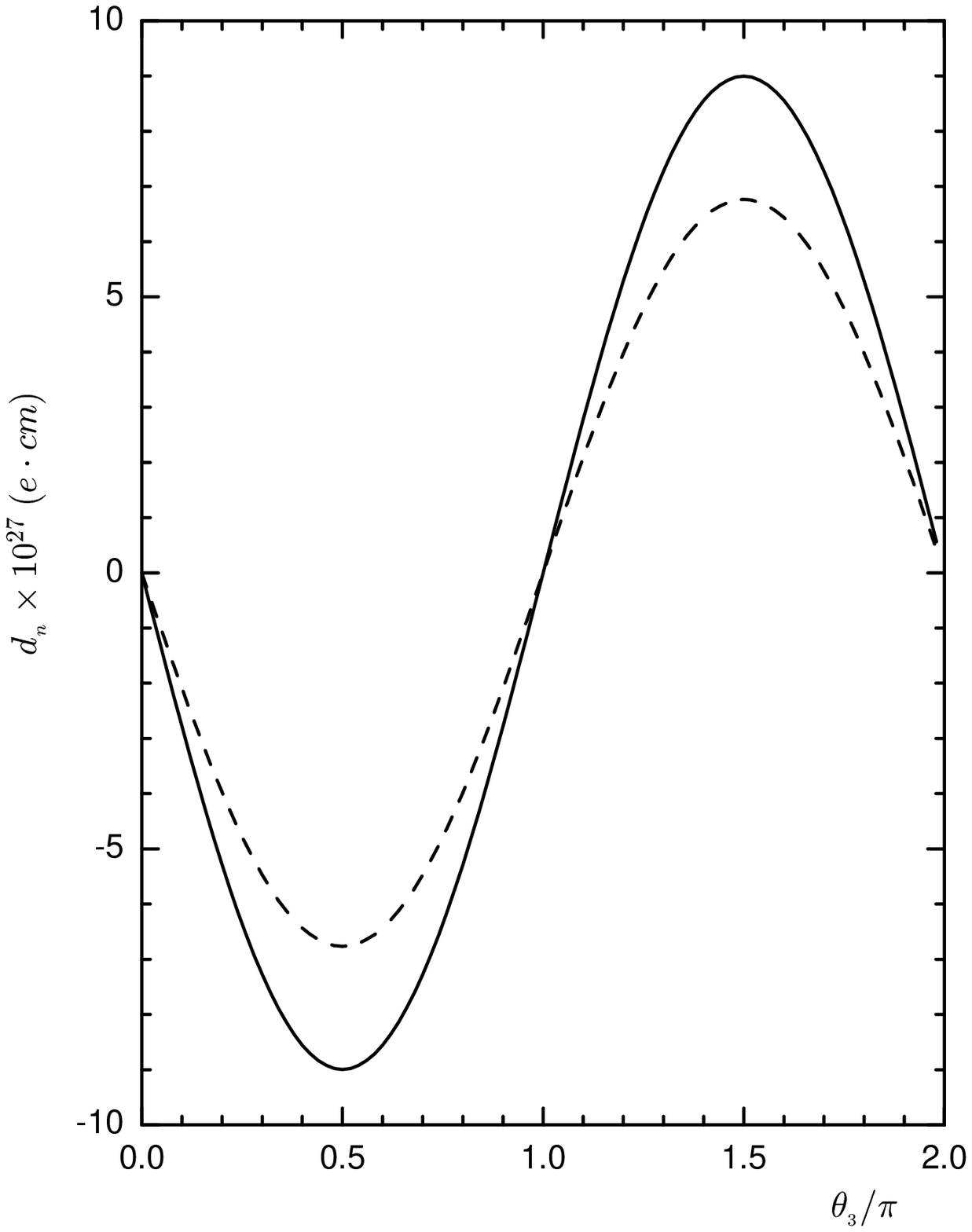}}
\end{picture}
\caption[]{Taking $\theta_{_1}=\theta_{_2}=0,\;m_{_{\tilde{Q}_1}}
=m_{_{\tilde U}}=m_{_{\tilde D}}=2.3\;{\rm TeV}$ and $\tan\beta=6$,
the neutron EDM varies with the $CP$ phase $\theta_{_3}$.}
\label{fig11}
\end{center}
\end{figure}

By including all the contributions present in Eq. (\ref{eq15}), we now
determine in our theoretical framework the variation of  the neutron EDM as a 
function of the CP phase $\theta_{_3}$. Taking
$\theta_{_1}=\theta_{_2}=0,\;m_{_{\tilde{Q}_1}} =m_{_{\tilde
U}}=m_{_{\tilde D}}=2.3\;{\rm TeV}$ and $\tan\beta=60$, we plot in Fig.
\ref{fig10} the neutron EDM versus the CP phase $\theta_{_3}$ , 
where the dashed line is the sum of the one-loop
results and the  contributions of the Weinberg operator, and the solid
line represents the results including two-loop gluino corrections
to the quark dipole moment operators. Fig. \ref{fig11} gives the corresponding 
plot for the case $\tan\beta=6$. It is clear that the
two-loop gluino corrections can reach the $30\%$ level for our choice
of the parameters.

In this work, we do not consider the effects of the CP phases associated with
the trilinear soft squark Yukawa coupling. This is reflected in our 
numerical results as a relatively strong decrease of the 
neutron EDM as a function increasing squark
masses. Moreover, for the model we employ here, the mass of the
lightest Higgs boson sets a strong constraint on the parameter
space of the new physics. As indicated in the literature 
\cite{Pilaftsis}, the CP violation
would cause changes to the neutral-Higgs-quark coupling, neutral
Higgs-gauge-boson coupling and self-coupling of Higgs boson. The
present experimental lower bound for the mass of the lightest Higgs boson is
relaxed to 60 GeV. In our numerical analysis we have
taken this constraint for the parameter space into account.

\section{The conclusion \label{sec4}}
\indent\indent We have investigated in this work  the corrections induced by
some two-loop gluino diagrams to the neutron EDM. Except a loop
factor, the two-loop contributions do not involve other
suppression factors. Since the dependence of the two-loop
corrections on the relevant CP violating phases differs from
that of the one-loop corrections, there is a region in the parameter space, where 
the two-loop corrections are comparable with the one-loop
corrections. Certainly, the present experimental upper bound on
the neutron EDM would set a very rigorous constraint on the possible values of 
CP phases if the  masses of new particles are  in the
TeV range. In order to circumvent this constraint, one can assume, as was done in 
the corresponding analysis at one-loop level, 
that the  squarks of the first two generations are heavy, {\it
i.e.} $\ge2 \;{\rm TeV}$, or invoke a mechanism that
causes an effective cancellation  among different contributions to
the quark EDMs.

Because the exact analytical expressions for the two-loop vacuum
integrals are derivable, one can extract, at least  in principle,  the Wilson
coefficients of the dipole moment operators from any two-loop
triangle diagrams and thereby find their contributions to the 
fermion anomalous magnetic dipole moments and EDMs. We will 
address our analysis to this issue in our coming
works.

\begin{acknowledgments}

The work has been supported by the Academy of Finland under the
contracts no.\ 104915 and 107293, and partly also by the National
Natural Sceince Foundation of China.
\end{acknowledgments}
\vspace{2.0cm}
\appendix
\section{The form factors\label{ap1}}
\begin{eqnarray}
&&{\cal N}_{_{\tilde{g}(1)}}^\gamma=-{4\over D}{q_1^2q_1\cdot
(q_2-q_1)\over(q_1^2-m_{_{\tilde{q}_j}}^2)^2}+{q_1\cdot(q_2-q_1)\over
q_1^2-m_{_{\tilde{q}_j}}^2}
-{4\over D}{q_1\cdot(q_2-q_1)q_2^2\over(q_2^2-|m_{_{\tilde g}}|^2)^2}
\nonumber\\
&&\hspace{1.4cm}
+{q_1\cdot(q_2-q_1)\over q_2^2-|m_{_{\tilde g}}|^2}
+{4\over(q_1^2-m_{_{\tilde{q}_j}}^2)
(q_2^2-|m_{_{\tilde g}}|^2)}\Big[{q_1^2q_1\cdot q_2\over D}
\nonumber\\
&&\hspace{1.4cm}
-{D(q_1\cdot q_2)^2-q_1^2q_2^2\over D(D-1)}\Big]
+{D-2\over D}{q_1\cdot(q_2-q_1)\over(q_2-q_1)^2-m_{_q}^2}
\;,\nonumber\\
&&{\cal N}_{_{\tilde{g}(2)}}^\gamma={2\over D}
{q_2\cdot(q_2-q_1)\over q_2^2-m_{_{\tilde{q}_i}}^2}+{4\over D}
{q_1^2q_2\cdot(q_2-q_1)\over(q_1^2-m_{_{\tilde{q}_j}}^2)^2}
-{q_2\cdot(q_2-q_1)\over q_1^2-m_{_{\tilde{q}_j}}^2}
\nonumber\\
&&\hspace{1.4cm}
+{4\over(q_1^2-m_{_{\tilde{q}_j}}^2)(q_2^2-|m_{_{\tilde g}}|^2)}
\Big[{q_1\cdot q_2q_2^2\over D}
-{D(q_1\cdot q_2)^2-q_1^2q_2^2\over D(D-1)}\Big]
\nonumber\\
&&\hspace{1.4cm}
+{4\over D}{q_2^2q_2\cdot(q_2-q_1)\over(q_2^2-|m_{_{\tilde g}}|^2)^2}
-{2+D\over D}{q_2\cdot(q_2-q_1)\over q_2^2-|m_{_{\tilde g}}|^2}
+{2-D\over D}{q_2\cdot(q_2-q_1)\over(q_2-q_1)^2-m_{_q}^2}
\;,\nonumber\\
&&{\cal N}_{_{\tilde{g}(1)}}^g=-{4\over D}{q_1^2q_1\cdot
(q_2-q_1)\over(q_1^2-m_{_{\tilde{q}_j}}^2)^2}+{17\over8}{q_1\cdot(q_2-q_1)\over
q_1^2-m_{_{\tilde{q}_j}}^2}-{4\over D}{q_1\cdot(q_2-q_1)q_2^2\over
(q_2^2-|m_{_{\tilde g}}|^2)^2}
\nonumber\\
&&\hspace{1.4cm}
+{4\over(q_1^2-m_{_{\tilde{q}_j}}^2)
(q_2^2-|m_{_{\tilde g}}|^2)}\Big[{q_1^2q_1\cdot q_2\over D}
-{D(q_1\cdot q_2)^2-q_1^2q_2^2\over D(D-1)}\Big]
\nonumber\\
&&\hspace{1.4cm}
-{1\over8}{q_1\cdot(q_2-q_1)\over q_2^2-|m_{_{\tilde g}}|^2}
-{18-9D\over8D}\Big[{q_2\cdot(q_2-q_1)\over q_2^2-|m_{_{\tilde g}}|^2}
+{q_2\cdot(q_2-q_1)\over q_2^2-m_{_{\tilde{q}_i}}^2}\Big]
\nonumber\\
&&\hspace{1.4cm}
+{10\over8}{2-D\over D}{q_1\cdot(q_2-q_1)\over(q_2-q_1)^2-m_{_q}^2}
\;,\nonumber\\
&&{\cal N}_{_{\tilde{g}(2)}}^g={4\over D}
{q_1^2q_2\cdot(q_2-q_1)\over(q_1^2-m_{_{\tilde{q}_j}}^2)^2}
+{4\over D}{q_2^2q_2\cdot(q_2-q_1)\over(q_2^2-|m_{_{\tilde g}}|^2)^2}
-{q_2\cdot(q_2-q_1)\over q_1^2-m_{_{\tilde{q}_j}}^2}
\nonumber\\
&&\hspace{1.4cm}
-{34-D\over8D}{q_2\cdot(q_2-q_1)\over q_2^2-|m_{_{\tilde g}}|^2}
+{4\over(q_1^2-m_{_{\tilde{q}_j}}^2)(q_2^2-|m_{_{\tilde g}}|^2)}
\Big[{q_1\cdot q_2q_2^2\over D}
\nonumber\\
&&\hspace{1.4cm}
-{D(q_1\cdot q_2)^2-q_1^2q_2^2\over D(D-1)}\Big]
-{1\over 4D}{q_2\cdot(q_2-q_1)\over q_2^2-m_{_{\tilde{q}_i}}^2}
+{9\over8}{q_2\cdot(q_2-q_1)\over q_1^2-|m_{_{\tilde g}}|^2}
\nonumber\\
&&\hspace{1.4cm}
-{10\over8}{2-D\over D}{q_2\cdot(q_2-q_1)\over(q_2-q_1)^2-m_{_q}^2}
\;,\nonumber\\
&&{\cal N}_{_{\chi_{_k}^0(1)}}^{g(a)}=
-{4\over D}{q_1^2q_1\cdot(q_2-q_1)\over(q_1^2-m_{_{\tilde{q}_j}}^2)^2}
+{q_1\cdot(q_2-q_1)\over q_1^2-m_{_{\tilde{q}_j}}^2}
-{4\over D}{q_1\cdot(q_2-q_1)q_2^2\over(q_2^2-|m_{_{\tilde g}}|^2)^2}
\nonumber\\
&&\hspace{1.4cm}
-{1\over8}{q_1\cdot(q_2-q_1)\over q_2^2-|m_{_{\tilde g}}|^2}
+{4\over(q_1^2-m_{_{\tilde{q}_j}}^2)
(q_2^2-|m_{_{\tilde g}}|^2)}\Big[{q_1^2q_1\cdot q_2\over D}
\nonumber\\
&&\hspace{1.4cm}
-{D(q_1\cdot q_2)^2-q_1^2q_2^2\over D(D-1)}\Big]
+{1\over8}{2-D\over D}{q_1\cdot(q_2-q_1)\over(q_2-q_1)^2-m_{_q}^2}
\;,\nonumber\\
&&{\cal N}_{_{\chi_{_k}^0(2)}}^{g(a)}={4\over D}
{q_1^2q_2\cdot(q_2-q_1)\over(q_1^2-m_{_{\tilde{q}_j}}^2)^2}
+{4\over D}{q_2^2q_2\cdot(q_2-q_1)\over(q_2^2-|m_{_{\tilde g}}|^2)^2}
-{q_2\cdot(q_2-q_1)\over q_1^2-m_{_{\tilde{q}_j}}^2}
\nonumber\\
&&\hspace{1.4cm}
-{34-D\over8D}{q_2\cdot(q_2-q_1)\over q_2^2-|m_{_{\tilde g}}|^2}
+{4\over(q_1^2-m_{_{\tilde{q}_j}}^2)(q_2^2-|m_{_{\tilde g}}|^2)}
\Big[{q_1\cdot q_2q_2^2\over D}
\nonumber\\
&&\hspace{1.4cm}
-{D(q_1\cdot q_2)^2-q_1^2q_2^2\over D(D-1)}\Big]
-{1\over 4D}{q_2\cdot(q_2-q_1)\over q_2^2-m_{_{\tilde{q}_i}}^2}
-{1\over8}{2-D\over D}{q_2\cdot(q_2-q_1)\over(q_2-q_1)^2-m_{_q}^2}
\;,\nonumber\\
&&{\cal N}_{_{\chi_{_k}^0(1)}}^{g(b)}=-{4\over D}{q_1^2q_1\cdot
(q_2-q_1)\over(q_1^2-m_{_{\tilde{q}_j}}^2)^2}+{17\over8}
{q_1\cdot(q_2-q_1)\over q_1^2-m_{_{\tilde{q}_j}}^2}
-{4\over D}{q_1\cdot(q_2-q_1)q_2^2\over(q_2^2-m_{_{\chi_{_k}^0}}^2)^2}
\nonumber\\
&&\hspace{1.4cm}
+{q_1\cdot(q_2-q_1)\over q_2^2-m_{_{\chi_{_k}^0}}^2}
+{4\over(q_1^2-m_{_{\tilde{q}_j}}^2)
(q_2^2-m_{_{\chi_{_k}^0}}^2)}\Big[{q_1^2q_1\cdot q_2\over D}
\nonumber\\
&&\hspace{1.4cm}
-{D(q_1\cdot q_2)^2-q_1^2q_2^2\over D(D-1)}\Big]
+{1\over8}{2-D\over D}{q_1\cdot(q_2-q_1)\over(q_2-q_1)^2-m_{_q}^2}
\nonumber\\
&&\hspace{1.4cm}
-{18-9D\over8D}\Big[{q_2\cdot(q_2-q_1)\over q_2^2-m_{_{\chi_{_k}^0}}^2}
+{q_2\cdot(q_2-q_1)\over q_2^2-m_{_{\tilde{q}_i}}^2}\Big]
\;,\nonumber\\
&&{\cal N}_{_{\chi_{_k}^0(2)}}^{g(b)}={4\over D}
{q_1^2q_2\cdot(q_2-q_1)\over(q_1^2-m_{_{\tilde{q}_j}}^2)^2}
+{4\over D}{q_2^2q_2\cdot(q_2-q_1)\over(q_2^2-m_{_{\chi_{_k}^0}}^2)^2}
-{q_2\cdot(q_2-q_1)\over q_1^2-m_{_{\tilde{q}_j}}^2}
\nonumber\\
&&\hspace{1.4cm}
-{2+D\over D}{q_2\cdot(q_2-q_1)\over q_2^2-m_{_{\chi_{_k}^0}}^2}
+{4\over(q_1^2-m_{_{\tilde{q}_j}}^2)(q_2^2-m_{_{\chi_{_k}^0}}^2)}
\Big[{q_1\cdot q_2q_2^2\over D}
\nonumber\\
&&\hspace{1.4cm}
-{D(q_1\cdot q_2)^2-q_1^2q_2^2\over D(D-1)}\Big]
+{2\over D}{q_2\cdot(q_2-q_1)\over q_2^2-m_{_{\tilde{q}_i}}^2}
+{9\over8}{q_2\cdot(q_2-q_1)\over q_1^2-|m_{_{\tilde g}}|^2}
\nonumber\\
&&\hspace{1.4cm}
-{1\over8}{2-D\over D}{q_2\cdot(q_2-q_1)\over(q_2-q_1)^2-m_{_q}^2}
\;.
\label{aeq1}
\end{eqnarray}

\section{The expression of $F_i(x_0, x_1, x_2, x_3, x_4)\;(i=1,\;2,
\;\cdots,\;7)$ \label{ap2}}
\begin{eqnarray}
&&F_1(x_0,x_1,x_2,x_3,x_4)=\Big\{\Xi_{_A}+\Xi_{_B}+\Xi_{_C}
+\Xi_{_D}+\Xi_{_E}+{1\over2}\Xi_{_F}\Big\}(x_0,x_1,x_2,x_3,x_4)\;,
\nonumber\\
&&F_2(x_0,x_1,x_2,x_3,x_4)=\Big\{\Xi_{_A}+{3\over2}\Xi_{_B}+\Xi_{_C}
+\Xi_{_D}+\Xi_{_E}+{1\over2}\Xi_{_F}\Big\}(x_0,x_2,x_1,x_4,x_3)
\nonumber\\
&&\hspace{3.8cm}
-{1\over2}\Xi_{_B}(x_0,x_4,x_1,x_2,x_3)\;,
\nonumber\\
&&F_3(x_0,x_1,x_2,x_3,x_4)=\Big\{\Xi_{_A}+{17\over8}\Xi_{_B}+\Xi_{_C}
-{1\over8}\Xi_{_D}+\Xi_{_E}-{5\over8}\Xi_{_F}\Big\}(x_0,x_1,x_2,x_3,x_4)
\nonumber\\
&&\hspace{3.8cm}
-{9\over16}\Big[\Xi_{_B}(x_0,x_2,x_1,x_4,x_3)
+\Xi_{_B}(x_0,x_4,x_1,x_2,x_3)\Big]\;,
\nonumber\\
&&F_4(x_0,x_1,x_2,x_3,x_4)=\Big\{\Xi_{_A}+\Xi_{_B}+\Xi_{_C}
-{1\over8}\Xi_{_D}+\Xi_{_E}-{1\over16}\Xi_{_F}\Big\}(x_0,x_1,x_2,x_3,x_4)\;,
\nonumber\\
&&F_5(x_0,x_1,x_2,x_3,x_4)=\Big\{\Xi_{_A}+{15\over16}\Xi_{_B}+\Xi_{_C}
+\Xi_{_D}+\Xi_{_E}-{1\over16}\Xi_{_F}\Big\}(x_0,x_2,x_1,x_4,x_3)
\nonumber\\
&&\hspace{3.8cm}
+{1\over16}\Xi_{_B}(x_0,x_4,x_1,x_2,x_3)\;,
\nonumber\\
&&F_6(x_0,x_1,x_2,x_3,x_4)=\Big\{e_{_q}\Big[\Xi_{_A}+\Xi_{_C}
+\Xi_{_D}+\Xi_{_E}\Big]+(2e_{_q}-e_{_Q})\Xi_{_B}
\nonumber\\
&&\hspace{3.8cm}
+{e_{_Q}\over2}\Xi_{_F}
\Big\}(x_0,x_1,x_2,x_3,x_4)
-{e_{_q}-e_{_Q}\over2}\Big[\Xi_{_B}(x_0,x_2,x_1,x_4,x_3)
\nonumber\\
&&\hspace{3.8cm}
+\Xi_{_B}(x_0,x_4,x_1,x_2,x_3)\Big]\;,
\nonumber\\
&&F_7(x_0,x_1,x_2,x_3,x_4)=\Big\{e_{_q}\Big[\Xi_{_A}+{3\over2}
\Xi_{_B}+\Xi_{_C}+\Xi_{_D}+\Xi_{_E}\Big]+{e_{_Q}\over2}
\Xi_{_F}\Big\}(x_0,x_2,x_1,x_4,x_3)
\nonumber\\
&&\hspace{3.8cm}
-{e_{_q}\over2}\Xi_{_B}(x_0,x_4,x_1,x_2,x_3)
-(e_{_q}-e_{_Q})\Xi_{_D}(x_0,x_2,x_3,x_4,x_1)\;.
\label{aeq2}
\end{eqnarray}

Where the functions $\Xi_{_\alpha}(x_0,x_1,x_2,x_3,x_4)\;(
\alpha=A,\;B,\;C,\;D,\;E,\;F)$ are formulated as
\begin{eqnarray}
&&\Xi_{_A}(x_0,x_1,x_2,x_3,x_4)=
-{1\over4}\Big\{\Big[A_1-A_0-{1\over2}
+{x_2\ln x_2-x_4\ln x_4\over x_2-x_4}\Big]
\nonumber\\
&&\hspace{1.0cm}\times
\Big[{3+2\ln x_1\over x_1-x_3}-{2x_1(1+2\ln x_1)\over(x_1-x_3)^2}
+{2(x_1^2\ln x_1-x_3^2\ln x_3)\over(x_1-x_3)^2}\Big]
\nonumber\\
&&\hspace{1.0cm}
+{x_3(x_0-x_2+x_3)\over(x_1-x_3)^3(x_2-x_4)}\Phi(x_0,x_1,x_2)
-{x_3(x_0+x_3-x_4)\over(x_1-x_3)^3(x_2-x_4)}\Phi(x_0,x_1,x_4)
\nonumber\\
&&\hspace{1.0cm}
+{x_1^2-x_0x_3-2x_1x_3+x_2x_3\over(x_1-x_3)^2(x_2-x_4)}
{\partial\Phi\over\partial x_1}(x_0,x_1,x_2)
\nonumber\\
&&\hspace{1.0cm}
-{x_1^2-x_0x_3-2x_1x_3+x_3x_4\over(x_1-x_3)^2(x_2-x_4)}
{\partial\Phi\over\partial x_1}(x_0,x_1,x_4)
\nonumber\\
&&\hspace{1.0cm}
+{x_1(x_0+x_1-x_2)\over2(x_1-x_3)(x_2-x_4)}
{\partial^2\Phi\over\partial^2x_1}(x_0,x_1,x_2)
\nonumber\\
&&\hspace{1.0cm}
-{x_1(x_0+x_1-x_4)\over2(x_1-x_3)(x_2-x_4)}{\partial^2\Phi
\over\partial^2x_1}(x_0,x_1,x_4)
\nonumber\\
&&\hspace{1.0cm}
-{x_3(x_0-x_2+x_3)\over(x_1-x_3)^3(x_2-x_4)}\Phi(x_0,x_3,x_2)
\nonumber\\
&&\hspace{1.0cm}
+{x_3(x_0+x_3-x_4)\over(x_1-x_3)^3(x_2-x_4)}\Phi(x_0,x_3,x_4)
\Big\}\;,
\label{aeq3}
\end{eqnarray}

\begin{eqnarray}
&&\Xi_{_B}(x_0,x_1,x_2,x_3,x_4)=
{1\over2}\Big\{\Big[A_1-A_0+{x_2\ln x_2
-x_4\ln x_4\over x_2-x_4}\Big]\cdot\Big[{1+\ln x_1\over x_1-x_3}
\nonumber\\
&&\hspace{1.0cm}
-{x_1\ln x_1-x_3\ln x_3\over(x_1-x_3)^2}\Big]
-\Big[{2\ln x_1+\ln^2x_1\over2(x_1-x_3)}-{x_1\ln^2x_1-x_3\ln^2x_3
\over2(x_1-x_3)^2}\Big]
\nonumber\\
&&\hspace{1.0cm}
-{x_0-x_2+x_3\over2(x_1-x_3)^2(x_2-x_4)}\Phi(x_0,x_1,x_2)
+{x_0+x_1-x_2\over2(x_1-x_3)(x_2-x_4)}{\partial\Phi\over\partial x_1}
(x_0,x_1,x_2)
\nonumber\\
&&\hspace{1.0cm}
+{x_0+x_3-x_4\over2(x_1-x_3)^2(x_2-x_4)}\Phi(x_0,x_1,x_4)
-{x_0+x_1-x_4\over2(x_1-x_3)(x_2-x_4)}{\partial\Phi\over\partial x_1}
(x_0,x_1,x_4)
\nonumber\\
&&\hspace{1.0cm}
+{x_0-x_2+x_3\over2(x_1-x_3)^2(x_2-x_4)}\Phi(x_0,x_3,x_2)
-{x_0+x_3-x_4\over2(x_1-x_3)^2(x_2-x_4)}\Phi(x_0,x_3,x_4)\Big\}\;,
\label{aeq4}
\end{eqnarray}

\begin{eqnarray}
&&\Xi_{_C}(x_0,x_1,x_2,x_3,x_4)=
{1\over4}\Big\{\Big[A_0+A_1+2\gamma_{_E}-2\ln(4\pi)+{1\over2}
-{x_1\ln x_1-x_3\ln x_3\over x_1-x_3}\Big]
\nonumber\\
&&\hspace{1.0cm}\times
\Big[{3+2\ln x_2\over x_2-x_4}-{2x_2(1+2\ln x_2)\over(x_2-x_4)^2}
+{2(x_2^2\ln x_2-x_4^2\ln x_4)\over(x_2-x_4)^3}\Big]
\nonumber\\
&&\hspace{1.0cm}
-{x_4(x_0+x_1-x_4)\over(x_1-x_3)(x_2-x_4)^3}\Phi(x_0,x_1,x_2)
+{x_0x_4+x_2^2-2x_2x_4+x_1x_4\over(x_1-x_3)(x_2-x_4)^2}{\partial\Phi
\over\partial x_2}(x_0,x_1,x_2)
\nonumber\\
&&\hspace{1.0cm}
-{x_2(x_0+x_1-x_2)\over2(x_1-x_3)(x_2-x_4)}{\partial^2\Phi
\over\partial^2 x_2}(x_0,x_1,x_2)
+{x_4(x_0+x_3-x_4)\over(x_1-x_3)(x_2-x_4)^3}\Phi(x_0,x_3,x_2)
\nonumber\\
&&\hspace{1.0cm}
-{x_0x_4+x_2^2-2x_2x_4+x_3x_4\over(x_1-x_3)(x_2-x_4)^2}{\partial\Phi
\over\partial x_2}(x_0,x_3,x_2)
+{x_2(x_0-x_2+x_3)\over2(x_1-x_3)(x_2-x_4)}{\partial^2\Phi
\over\partial^2 x_2}(x_0,x_3,x_2)
\nonumber\\
&&\hspace{1.0cm}
+{x_4(x_0+x_1-x_4)\over(x_1-x_3)(x_2-x_4)^3}\Phi(x_0,x_1,x_4)
-{x_4(x_0+x_3-x_4)\over(x_1-x_3)(x_2-x_4)^3}\Phi(x_0,x_3,x_4)
\Big\}\;,
\label{aeq5}
\end{eqnarray}

\begin{eqnarray}
&&\Xi_{_D}(x_0,x_1,x_2,x_3,x_4)=
{1\over2}\Big\{\Big[-2\gamma_{_E}-2\ln(4\pi)
-A_0-A_1+{x_1\ln x_1-x_3\ln x_3\over x_1-x_3}\Big]
\nonumber\\
&&\hspace{1.0cm}\times
\Big[{1+\ln x_2\over x_2-x_4}-{x_2\ln x_2-x_4\ln x_4\over(x_2-x_4)^2}\Big]
+\Big[{\ln^2x_2+2\ln x_2\over2(x_2-x_4)}
-{x_2\ln^2x_2-x_4\ln^2x_4\over2(x_2-x_4)^2}\Big]
\nonumber\\
&&\hspace{1.0cm}
-{x_0+x_1-x_4\over2(x_1-x_3)(x_2-x_4)^2}\Phi(x_0,x_1,x_2)
+{x_0+x_1-x_2\over2(x_1-x_3)(x_2-x_4)}{\partial\Phi\over
\partial x_2}(x_0,x_1,x_2)
\nonumber\\
&&\hspace{1.0cm}
+{x_0+x_1-x_4\over2(x_1-x_3)(x_2-x_4)^2}\Phi(x_0,x_1,x_4)
+{x_0+x_3-x_4\over2(x_1-x_3)(x_2-x_4)^2}\Phi(x_0,x_3,x_2)
\nonumber\\
&&\hspace{1.0cm}
-{x_0-x_2+x_3\over2(x_1-x_3)(x_2-x_4)}{\partial\Phi\over
\partial x_2}(x_0,x_3,x_2)
-{x_0+x_3-x_4\over2(x_1-x_3)(x_2-x_4)^2}\Phi(x_0,x_3,x_4)
\Big\}\;,
\label{aeq6}
\end{eqnarray}

\begin{eqnarray}
&&\Xi_{_E}(x_0,x_1,x_2,x_3,x_4)=
{1\over6}\Big\{(A_1-A_0)\cdot\Big[{1+\ln x_2\over x_2-x_4}
-{x_2\ln x_2-x_4\ln x_4\over(x_2-x_4)^2}
-{2+2\ln x_1\over x_1-x_3}
\nonumber\\
&&\hspace{1.0cm}
+{2(x_1\ln x_1-x_3\ln x_3)\over(x_1-x_3)^2}\Big]
+3\Big[{x_1+2x_1\ln x_1\over x_1-x_3}
-{x_1^2\ln x_1-x_3^2\ln x_3\over(x_1-x_3)^2}\Big]\cdot
\Big[{1+\ln x_2\over x_2-x_4}
\nonumber\\
&&\hspace{1.0cm}
-{x_2\ln x_2-x_4\ln x_4\over(x_2-x_4)^2}\Big]
-{3\over2}\Big[{x_2+2x_2\ln x_2\over x_2-x_4}
-{x_2^2\ln x_2-x_4^2\ln x_4\over(x_2-x_4)^2}\Big]
+{\ln x_1+2\ln x_1\over x_1-x_3}
\nonumber\\
&&\hspace{1.0cm}
-{(x_1\ln^2x_1-x_3\ln^2x_3)\over(x_1-x_3)^2}
+{\ln x_2+2\ln x_2\over x_2-x_4}-{(x_2\ln^2x_2-x_4\ln^2x_4)
\over(x_2-x_4)^2}
\nonumber\\
&&\hspace{1.0cm}
+{2[x_1^2x_4+x_2^2x_3+(x_0-2x_1-2x_2)x_3x_4]\over(x_1-x_3)^2(x_2-x_4)^2}
\ln x_1\ln x_2
\nonumber\\
&&\hspace{1.0cm}
-{2[x_1^2-x_2x_3-2x_1x_3+x_0x_3]\over(x_1-x_3)^2(x_2-x_4)}\ln x_1
-{2[x_2^2-x_1x_4-2x_2x_4+x_0x_4]\over(x_1-x_3)(x_2-x_4)^2}\ln x_2
\nonumber\\
&&\hspace{1.0cm}
-{2(x_1+x_2-x_0)\over(x_1-x_3)(x_2-x_4)}
-{2x_4[x_1^2+x_3(x_0-2x_1-x_4)]\over(x_1-x_3)^2(x_2-x_4)^2}
\ln x_1\ln x_4
\nonumber\\
&&\hspace{1.0cm}
-{2x_4(x_1+x_4-x_0)\over(x_1-x_3)(x_2-x_4)^2}\ln x_4
-{2x_3[x_2^2+x_4(x_0-2x_2-x_3)]\over(x_1-x_3)^2(x_2-x_4)^2}
\ln x_2\ln x_3
\nonumber\\
&&\hspace{1.0cm}
-{2x_3(x_2+x_3-x_0)\over(x_1-x_3)^2(x_2-x_4)}\ln x_3
-{2x_3x_4(x_3+x_4-x_0)\over(x_1-x_3)^2(x_2-x_4)^2}\ln x_3\ln x_4
\nonumber\\
&&\hspace{1.0cm}
+{\Phi(x_0,x_1,x_2)\over2(x_1-x_3)^2(x_2-x_4)^2}\Big[2x_0^2+x_1^2
-2x_2^2-x_0x_3-4x_0x_4-2x_1x_3+4x_2x_4+x_3x_4\Big]
\nonumber\\
&&\hspace{1.0cm}
+{x_1^2-2x_0^2+2x_2^2+x_0x_1+4x_0x_4-x_1x_4-4x_2x_4\over
2(x_1-x_3)(x_2-x_4)^2}{\partial\Phi\over\partial x_1}(x_0,x_1,x_2)
\nonumber\\
&&\hspace{1.0cm}
-{2x_0^2+x_1^2+2x_2^2-4x_0x_2-x_0x_3-2x_1x_3+x_2x_3\over
2(x_1-x_3)^2(x_2-x_4)}{\partial\Phi\over\partial x_2}(x_0,x_1,x_2)
\nonumber\\
&&\hspace{1.0cm}
+{2x_0^2-x_1^2+2x_2^2-x_0x_1-4x_0x_2-x_1x_2\over2(x_1-x_3)(x_2-x_4)}
{\partial^2\Phi\over\partial x_1\partial x_2}(x_0,x_1,x_2)
\nonumber\\
&&\hspace{1.0cm}
-{2x_0^2+x_1^2+2x_4^2-x_0x_3-4x_0x_4-2x_1x_3+x_3x_4\over
2(x_1-x_3)^2(x_2-x_4)^2}\Phi(x_0,x_1,x_4)
\nonumber\\
&&\hspace{1.0cm}
+{2x_0^2-x_1^2+2x_4^2-x_0x_1-4x_0x_4-x_1x_4\over2(x_1-x_3)(x_2-x_4)^2}
{\partial\Phi\over\partial x_1}(x_0,x_1,x_4)
\nonumber\\
&&\hspace{1.0cm}
-{2x_0^2-2x_2^2-x_3^2-x_0x_3-4x_0x_4+4x_2x_4+x_3x_4\over2(x_1-x_3)^2
(x_2-x_4)^2}\Phi(x_0,x_3,x_2)
\nonumber\\
&&\hspace{1.0cm}
+{2x_0^2+2x_2^2-x_3^2-4x_0x_2-x_0x_3-x_2x_3\over2(x_1-x_3)^2(x_2-x_4)}
{\partial\Phi\over\partial x_2}(x_0,x_3,x_2)
\nonumber\\
&&\hspace{1.0cm}
+{2x_0^2-x_3^2+2x_4^2-x_0x_3-4x_2x_4-x_3x_4\over2(x_1-x_3)^2(x_2-x_4)^2}
\Phi(x_0,x_3,x_4)\Big\}\;,
\label{aeq7}
\end{eqnarray}

\begin{eqnarray}
&&\Xi_{_F}(x_0,x_1,x_2,x_3,x_4)=
{1\over4}\Big\{{1\over(x_1-x_3)(x_2-x_4)}
\Big[\Phi(x_0,x_1,x_2)-\Phi(x_0,x_1,x_4)
\nonumber\\
&&\hspace{1.0cm}
-\Phi(x_0,x_3,x_2)+\Phi(x_0,x_3,x_4)\Big]
+{x_0+x_1-x_2\over(x_1-x_3)(x_2-x_4)}{\partial\Phi\over\partial x_0}
(x_0,x_1,x_2)
\nonumber\\
&&\hspace{1.0cm}
-{x_0+x_1-x_4\over(x_1-x_3)(x_2-x_4)}{\partial\Phi\over\partial x_0}
(x_0,x_1,x_4)
-{x_0+x_3-x_2\over(x_1-x_3)(x_2-x_4)}{\partial\Phi\over\partial x_0}
(x_0,x_3,x_2)
\nonumber\\
&&\hspace{1.0cm}
+{x_0+x_3-x_4\over(x_1-x_3)(x_2-x_4)}{\partial\Phi\over\partial x_0}
(x_0,x_3,x_4)\Big\}\;,
\label{aeq8}
\end{eqnarray}
with $A_0=-\gamma_{_E}+\ln(4\pi x_{_\mu}),\;A_1=3-2\gamma_{_E}+2\ln{4\pi
\over x_{_\mu}}$. Here, $x_{_\mu}=\mu_{_{\rm NP}}^2/m_{_{\rm w}}^2$,
and $\mu_{_{\rm NP}}$ is the scale to integrate heavy particles out.

Defining $\lambda^2=x_0^2+x_1^2+x_2^2-2x_0x_1-2x_0x_2-2x_1x_2$,
the two-loop vacuum function $\Phi(x_0,x_1,x_2)$ is written as
\begin{itemize}
\item $\lambda^2>0,\;\sqrt{x_1}+\sqrt{x_2}<\sqrt{x_0}$:
\begin{eqnarray}
&&\Phi(x_0,x_1,x_2)=(x_0+x_1-x_2)\ln x_0\ln x_1+(x_0-x_1+x_2)\ln x_0\ln x_2
\nonumber\\
&&\hspace{1.8cm}
+(-x_0+x_1+x_2)\ln x_1\ln x_2+\lambda\Big\{2\ln\Big({x_0+x_1-x_2-\lambda
\over2x_0}\Big)
\nonumber\\
&&\hspace{1.8cm}\times
\ln\Big({x_0-x_1+x_2-\lambda\over2x_0}\Big)-\ln{x_1\over x_0}
\ln{x_2\over x_0}-2L_{i_2}\Big({x_0+x_1-x_2-\lambda\over2x_0}\Big)
\nonumber\\
&&\hspace{1.8cm}
-2L_{i_2}\Big({x_0-x_1+x_2-\lambda\over2x_0}\Big)+{\pi^2\over3}\Big\}\;,
\label{aeq9}
\end{eqnarray}
where $L_{i_2}(x)$ is the spence function;

\item $\lambda^2>0,\;\sqrt{x_0}+\sqrt{x_2}<\sqrt{x_1}$:
\begin{eqnarray}
&&\Phi(x_0,x_1,x_2)={\rm Eq.}(\ref{aeq9})(x_0\leftrightarrow x_1)\;;
\label{aeq10}
\end{eqnarray}

\item $\lambda^2>0,\;\sqrt{x_0}+\sqrt{x_1}<\sqrt{x_2}$:
\begin{eqnarray}
&&\Phi(x_0,x_1,x_2)={\rm Eq.}(\ref{aeq9})(x_0\leftrightarrow x_2)\;;
\label{aeq11}
\end{eqnarray}

\item $\lambda^2<0$:
\begin{eqnarray}
&&\Phi(x_0,x_1,x_2)=(x_0+x_1-x_2)\ln x_0\ln x_1+(x_0-x_1+x_2)\ln x_0\ln x_2
\nonumber\\
&&\hspace{1.8cm}
+(-x_0+x_1+x_2)\ln x_1\ln x_2-2\sqrt{|\lambda^2|}\Big\{Cl_2\Big(2\arccos(
{-x_0+x_1+x_2\over2\sqrt{x_1x_2}})\Big)
\nonumber\\
&&\hspace{1.8cm}
+Cl_2\Big(2\arccos({x_0-x_1+x_2\over2\sqrt{x_0x_2}})\Big)+Cl_2\Big(2\arccos(
{x_0+x_1-x_2\over2\sqrt{x_0x_1}})\Big)\Big\}\;,
\label{aeq12}
\end{eqnarray}
where $Cl_2(x)$ denotes the Clausen function.
\end{itemize}

\end{document}